\documentclass[letterpaper, journal]{IEEEtran}
\usepackage{epsfig} 
\usepackage{varwidth}
\usepackage[abs]{overpic}
\usepackage{algorithmic}

\usepackage[caption=false,font=footnotesize]{subfig}
\newcommand{\squeezeup}{\vspace{-2.5mm}}
\newcounter{numberlatin}
\newcommand{\rom}[1]{\romannumeral #1}
\usepackage{cite}

\usepackage{enumerate}

\usepackage{color}

\ifCLASSINFOpdf
\else
\fi
\usepackage{amsmath}
\usepackage{amssymb}  

\usepackage{algorithmic}
\usepackage{caption}

\usepackage[caption=false,font=footnotesize]{subfig}
\begin{document}
\IEEEpubid{\copyright~2018 IEEE. Personal use is permitted. For any other purposes, permission must be obtained from the IEEE.  DOI: 10.1109/MED.2018.8442796 }
\IEEEpubidadjcol

%
\title{Geometric Surface-Based Tracking Control of a Quadrotor UAV
 }
%
%
%

\author{Michalis~Ramp$^1$~and~Evangelos~Papadopoulos$^2$,~\IEEEmembership{Senior~Member,~IEEE}
\thanks{$^1$M. Ramp is with the Department of Mechanical Engineering, National Technical University of Athens, (NTUA) 15780 Athens, Greece.
{\tt\small rampmich@mail.ntua.gr}}
\thanks{$^2$E. Papadopoulos is with the Department of Mechanical Engineering, NTUA, 15780 Athens (tel: +30-210-772-1440; fax: +30-210-772-1455).
{\tt\small egpapado@central.ntua.gr}}
}

\maketitle

\begin{abstract}
New quadrotor UAV control algorithms are developed, based on nonlinear surfaces composed of tracking errors that evolve directly on the nonlinear configuration manifold, thus inherently including in the control design the nonlinear characteristics of the SE(3) configuration space.
In particular, geometric surface-based controllers are developed and are shown, through rigorous stability proofs, to have desirable almost global closed loop properties.
For the first time in regards to the geometric literature, a region of attraction independent of the position error is identified and its effects are analyzed.
The effectiveness of the proposed `surface based' controllers are illustrated by simulations of aggressive maneuvers in the presence of disturbances and motor saturation.
\end{abstract}


%
\IEEEpeerreviewmaketitle

\section{Introduction}
%
%
%
%
Quadrotor unmanned aerial vehicles are characterized by a simple mechanical structure comprised of two pairs of counter rotating outrunner motors where each one is driving a dedicated propeller, resulting in a platform with high thrust-to-weight ratio, able to achieve vertical takeoff and landing (VTOL) maneuvers and operate in a broad spectrum of flight scenarios.
Quadrotors have good flight endurance characteristics and acceptable payload transporting potential for a plethora of applications. 
Although the quadrotor UAV has six degrees of freedom, it is underactuated since it has only four inputs and can only track four commands or less.

A plethora of theoretical and experimental works regarding quadrotors exist including results demonstrating aerobatic maneuvers \cite{geomquadlee}, decentralized collision avoidance for multiple quadrotors \cite{Gillula}, safe passage schemes satisfying constraints on velocities, accelerations, and inputs \cite{Mellinger}, backsteping control laws \cite{MahonyHamel}, and hybrid global/robust controllers \cite{Casau}, \cite{Abdessameud},\cite{NaldiRob}.

This work follows the geometric framework.
A geometric nonlinear control system (GNCS) for a quadrotor UAV is developed directly on the special Euclidean group, thus inherently entailing in the control design the characteristics of the nonlinear configuration manifold, and avoiding singularities and ambiguities associated with minimal attitude representations.
The key contributions of this work are: 
(a) An attitude and a position controller is developed based on nonlinear surfaces composed by tracking errors that evolve directly on the nonlinear configuration manifold.
These controllers allow for precision pose tracking by tuning three gains per controller and are able to follow an attitude tracking command and a position tracking command.
(b) In contrast to other GNCSs such as like \cite{geomquadlee}, \cite{geommac} -\cite{qeopidfar}, rigorous stability proofs are developed and regions of attraction both with and without restrictions on the initial position/velocity error are identified.
A region of attraction independent of the initial position/velocity error is desired since it introduces simplicity in trajectory design.
The proposed strategies are validated in simulation in the presence of motor saturation and wind disturbances.

\section{Quadrotor Kinetics Model}

\begin{figure}[h!]
\centering
\includegraphics[width=1\columnwidth]{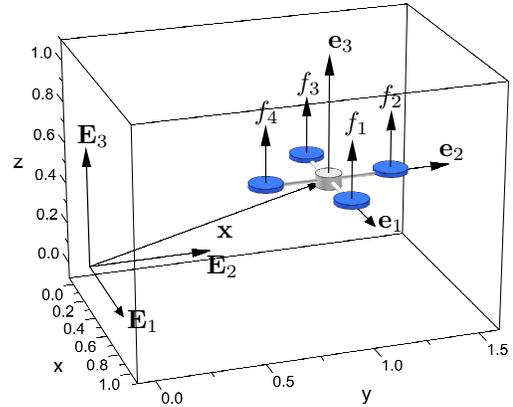}
\put(-82,74){\parbox{\columnwidth}{$\mathbf{e}_{1}$}}
\put(-59,101){\parbox{\columnwidth}{$\mathbf{e}_{2}$}}
\put(-101,142){\parbox{\columnwidth}{$\mathbf{e}_{3}$}}
\put(-143,71){\parbox{\columnwidth}{$\mathbf{x}$}}
\put(-177,37){\parbox{\columnwidth}{$\mathbf{E}_{1}$}}
\put(-147,58){\parbox{0.4\linewidth}{$\mathbf{E}_{2}$}}
\put(-196,107){\parbox{0.4\linewidth}{$\mathbf{E}_{3}$}}
\put(-95,112){\parbox{0.4\linewidth}{$f_{1}$}}
\put(-82,122){\parbox{0.4\linewidth}{$f_{2}$}}
\put(-113,127){\parbox{0.4\linewidth}{$f_{3}$}}
\put(-129,116){\parbox{0.4\linewidth}{$f_{4}$}}
\caption{Quadrotor with coordinate frames, and actuator forces.}
\label{Quadrotor}
\end{figure}

\IEEEpubidadjcol
The quadrotor studied is comprised by two pairs of counter rotating out-runner motors, see Fig. \ref{Quadrotor}.
Each motor drives a dedicated propeller and generates thrust and torque normal to the plane produced by the centers of mass (CM) of the four rotors. 
An inertial reference frame  I$_{R}\big\{\mathbf{E}_1,\mathbf{E}_2,\mathbf{E}_3\big\}$ and a body-fixed frame I$_{b}\big\{\mathbf{e}_1,\mathbf{e}_2,\mathbf{e}_3\big\}$ are employed with the origin of the latter to be located at the quadrotor CM, which belongs to the four rotor CM plane. 
Vectors $\mathbf{e}_{1}$ and $\mathbf{e}_{2}$ are co-linear with the two quadrotor legs , see Fig. \ref{Quadrotor}.

The following apply throughout the paper.
The actual control input is the thrust of each propeller, which is co-linear with $\mathbf{e}_{3}$.
The first and third propellers generate positive thrust when rotating clockwise, while the second and fourth propellers generate positive thrust when rotating counterclockwise.
The magnitude of the total thrust is denoted by $f=\sum_{i=1}^{4} f_i\in\mathbb{R}$, where $f_{i}$ and other system variables are defined in Table \ref{Table}.

\begin{table}[h]
\caption{Definitions of variables.}
\label{Table}
\begin{center}
\begin{tabular}{|l|l|}
\hline
$\mathbf{x}\in\mathbb{R}^3$ & Quadrotor CM position wrt. I$_R$ in I$_R$\\
\hline
$\mathbf{v}\in\mathbb{R}^3$ & Quadrotor CM velocity wrt. I$_R$ in I$_R$\\
\hline
$^{b}\boldsymbol{\omega}\in\mathbb{R}^3$ & Quadrotor angular velocity wrt I$_R$ in I$_{b}$\\
\hline
$\mathbf{R}\in\text{SO}\left(3\right)$ & Rotation matrix from $\mathbf{I}_b$ to $\mathbf{I}_R$ frame\\
\hline
$^b\mathbf{u}\in\mathbb{R}^3$ & Control moment $^b\mathbf{u}{=}[{}^{b}u_{1};{}^{b}u_{2};{}^{b}u_{3}]$ in I$_b$\\
\hline
$f_i\in\mathbb{R}$ & Force produced by the i-th propeller along $\mathbf{e}_{3}$ \\
\hline
$b_T\in\mathbb{R}^{+}$ & Torque coefficient \\
\hline
$g\in\mathbb{R}$ & Gravity constant\\
\hline
$d\in\mathbb{R}^{+}$ & Distance between system CM and each motor axis\\
\hline
$\mathbf{J}\in\mathbb{R}^{3\times3}$ & Inertial matrix (IM) of the quadrotor in I$_b$\\
\hline
$m\in\mathbb{R}$ & Quadrotor total mass\\
\hline
$\lambda_{min,max}(.)$ & Minimum, maximum eigenvalue of $(.)$ respectively\\
\hline
\end{tabular}
\end{center}
\end{table}

The motor torques, $\boldsymbol{\tau}_{i}$, corresponding to each propeller are assumed to be proportional to thrust,
\begin{IEEEeqnarray}{C}
\boldsymbol{\tau}_{i}=(-1)^{i}b_{T}f_{i}\mathbf{e}_{3},\; i=1,..,4
\end{IEEEeqnarray}
where the $(-1)^{i}$ term connects each propeller with the correct rotation direction (clockwise and counterclockwise).
The control inputs include the total propeller thrust $f$ and moment, $^{b}\mathbf{u}$, given by,
\begin{IEEEeqnarray}{C}
\begin{bmatrix}
f\\
^b\mathbf{u}
\end{bmatrix}
=
\begin{bmatrix}
1&1&1&1\\
0&d&0&-d\\
-d&0&d&0\\
-b_{T}&b_{T}&-b_{T}&b_{T}\\
\end{bmatrix}\!\!\mathbf{F},\;
\mathbf{F}=\begin{bmatrix}
f_{1}\\
f_{2}\\
f_{3}\\
f_{4}\\
\end{bmatrix}\IEEEeqnarraynumspace
\label{eq:distribution}
\end{IEEEeqnarray}
with $\mathbf{F}\in\mathbb{R}^{4}$ the thrust vector, and the $4\times4$ matrix to be always full rank for $d,b_{T}\in\mathbb{R}^{+}$.

The spatial configuration of the quadrotor UAV is described by the quadrotor attitude and the location of its center of mass, both with respect to $\mathbf{I}_{R}$.
The configuration manifold is the special Euclidean group SE(3)=$\mathbb{R}^{3}\times\text{SO(3)}$.
The total thrust produced by the propellers, in $\mathbf{I}_{R}$, is given by $\mathbf{R}f\mathbf{e}_{3}$.
The equations of motion of the quadrotor are given by,
\begin{IEEEeqnarray}{rCl}
\mathbf{\dot{x}}&=&{}\mathbf{v}\IEEEnonumber\IEEEeqnarraynumspace\\
m\dot{\mathbf{v}}&=&-mg\mathbf{E}_{3}+\mathbf{R}f\mathbf{e}_{3}+\boldsymbol{\delta}_{x}\IEEEyesnumber\label{eq:position}\\
\mathbf{J}{}^{b}\dot{\boldsymbol{\omega}}&=&{}^{b}\mathbf{u}-{}^{b}\boldsymbol{\omega}\times\mathbf{J}{}^{b}\boldsymbol{\omega}+\boldsymbol\delta_{R}\IEEEyesnumber\IEEEeqnarraynumspace\label{eq:attitude}\\
\dot{\mathbf{R}}&=&\mathbf{R}S({}^{b}\boldsymbol{\omega}) \IEEEyesnumber
\label{attitude_kinem}\end{IEEEeqnarray}
where $\boldsymbol\delta_{x}$, $\boldsymbol\delta_{R}$ are disturbance terms and $S(.):\mathbb{R}^{3}\rightarrow\mathfrak{so}(3)$ is the cross product map given by,
\begin{IEEEeqnarray}{rCl}
\begin{array}{c}
S(\mathbf{r}){=}[{0},{-r_{3}},{r_{2}};{r_{3}},{0},{-r_{1}};{-r_{2}},{r_{1}},0]\\
{S^{-1}}(S(\mathbf{r})){=}\mathbf{r}
\end{array}\label{iso}
\end{IEEEeqnarray}

\section{Quadrotor Tracking Controls\label{tracControls}}
Given the underactuated nature of quadrotors, in this paper two flight modes are considered:
\begin{itemize}
\item \textit{Attitude Control Mode}:
The controller achieves tracking for the attitude of the quadrotor UAV.
\item \textit{Position Control Mode}:
The controller achieves tracking for the quadrotor CM position and a pointing attitude associated with the quadrotor yaw.
\end{itemize}

Using these flight modes in suitable successions, a quadrotor can perform a complex desired flight maneuver.
Moreover it will be shown that each mode has stability properties that allow the safe switching between flight modes (end of Section \ref{tracControls}).

\newcounter{Prop1}
\addtocounter{Prop1}{1}
\newcounter{Prop2}
\addtocounter{Prop2}{2}
\newcounter{Prop3}
\addtocounter{Prop3}{3}
\newcounter{Prop4}
\addtocounter{Prop4}{4}
\newcounter{Prop5}
\addtocounter{Prop5}{5}
\newcounter{sub}
\addtocounter{sub}{1}

\subsection{Attitude Control Mode (ACM)\label{conmodatt}}
An attitude control system able to follow an arbitrary smooth desired orientation $\mathbf{R}_{d}(t)\in\text{SO(3)}$ and its associated angular velocity $^{b}\boldsymbol{\omega}_{d}(t)\in\mathbb{R}^{3}$ is developed next under the assumption that $\boldsymbol{\delta}_{R}=0_{3\times1}$.
\subsubsection{Attitude tracking errors}
For a given tracking command ($\mathbf{R}_{d}$, $^{b}\boldsymbol{\omega}_{d}$) and current attitude and angular velocity ($\mathbf{R}$, $^{b}\boldsymbol{\omega}$), two sets of geometric attitude tracking errors are considered.
Each set consists of an \textit{attitude error function} $\Psi:\text{SO(3)}\times\text{SO(3)}\rightarrow\mathbb{R}$, and an \textit{attitude error vector} $\mathbf{e}_{R}\in\mathbb{R}^{3}$, defined as follows.
The first set is, \cite{qeoadapclee}:
\begin{IEEEeqnarray}{rCl}
\Psi(\mathbf{R},\mathbf{R}_{d})&=&\frac{1}{2}tr[\mathbf{I}-\mathbf{R}^{T}_{d}\mathbf{R}]\geq 0
\label{error_function}\\
\mathbf{e}_{R}(\mathbf{R},\mathbf{R}_{d})&=&\frac{1}{2}S^{-1}(\mathbf{R}^{T}_{d}\mathbf{R}-\mathbf{R}^{T}\mathbf{R}_{d})
\label{att_error}
\end{IEEEeqnarray}
where $tr[.]$ is the trace function. 
The second according to \cite{err_fun}:
\begin{IEEEeqnarray}{rCl}
\Psi(\mathbf{R},\mathbf{R}_{d})&=&2-\sqrt{1+tr[\mathbf{R}^{T}_{d}\mathbf{R}]}\geq 0
\label{error_function_A}\\
\mathbf{e}_{R}(\mathbf{R},\mathbf{R}_{d})&=&\frac{1}{2}S^{-1}(\mathbf{R}^{T}_{d}\mathbf{R}{-}\mathbf{R}^{T}\mathbf{R}_{d})(1{+}tr[\mathbf{R}^{T}_{d}\mathbf{R}])^{-\frac{1}{2}}
\label{att_error_A}
\end{IEEEeqnarray}
Both (\ref{error_function}),(\ref{error_function_A}) yield the angular velocity error vector, $\mathbf{e}_{\omega}{\in}\mathbb{R}^{3}$,
\begin{IEEEeqnarray}{rCl}
\mathbf{e}_{\omega}(\mathbf{R},{}^{b}\boldsymbol{\omega},\mathbf{R}_{d},{}^{b}\boldsymbol{\omega}_{d})&=&{}^{b}\boldsymbol{\omega}-\mathbf{R}^{T}\mathbf{R}_{d}{}^{b}\boldsymbol{\omega}_{d}
\label{ang_vel_error}
\end{IEEEeqnarray}
For the ACM, the controller is designed to be compatible with both sets of $\mathbf{e}_{R}$.
This is because the first set given by $\{(\ref{error_function}),(\ref{att_error})\}$ bestows excellent tracking properties to the controller if the orientation tracking error remains less than $90^{o}$ wrt. an axis-angle rotation; however for an orientation error larger than $90^{o}$, the magnitude of the attitude error vector, (\ref{att_error}), is not proportional to the orientation error and results to deteriorating performance as the state approaches the antipodal equilibrium (see \cite{err_fun} for more details).
In contrast to this, the second set given by $\{(\ref{error_function_A}),(\ref{att_error_A})\}$ does not suffer from this problem but is marginally outperformed by the first set if the attitude error is less than $90^{o}$.
Thus depending on the flight conditions, the user can choose which set of attitude tracking errors to use.

Note that the maximum attitude difference, that of 180$^{o}$ with respect to an equivalent axis-angle rotation between $\mathbf{R}$ and $\mathbf{R}_{d}$, occurs when the rotation matrices are antipodal; then (\ref{error_function}) or (\ref{error_function_A}) yield $\Psi(\mathbf{R},\mathbf{R}_{d})$=2, i.e. 100\% error.
If both rotation matrices express the same attitude i.e., $\mathbf{R}$=$\mathbf{R}_{d}$, then $\Psi(\mathbf{R},\mathbf{R}_{d})$=0, i.e. 0\% error.
Important properties regarding (\ref{error_function})-(\ref{ang_vel_error}), including the associated attitude error dynamics used throughout this work are included in \text{Proposition \arabic{Prop1}} and \text{Proposition \arabic{Prop2}}  found in Appendix \ref{appA}.

\subsubsection{Attitude tracking controller}
A controller is developed stabilizing $\mathbf{e}_{R}$, $\mathbf{e}_{\omega}$, to zero exponentially, almost globally under the assumption that $\boldsymbol{\delta}_{R}=0_{3\times1}$.

\textbf{Proposition \arabic{Prop3}.} 
For $\eta,k_{R},k_{\omega}\in\mathbb{R}^{+}$, with,
\begin{IEEEeqnarray}{C}
\eta>{k_{R}}/{k_{\omega}}^{2}\label{eq:hta}
\end{IEEEeqnarray}
and initial conditions satisfying,
\begin{IEEEeqnarray}{C}
\Psi(\mathbf{R}(0),\mathbf{R}_{d}(0))<2
\label{Psi_0}\\
\lVert\mathbf{e}_{\omega}(0)\rVert^{2}<2\eta k_{R}\left(2-\Psi(\mathbf{R}(0),\mathbf{R}_{d}(0))\right)\label{surface_0}
\end{IEEEeqnarray}
and for a desired arbitrary smooth attitude $\mathbf{R}_{d}(t)\in\text{SO(3)}$ in,
\begin{IEEEeqnarray}{rCl}
L_{2}&=&\{(\mathbf{R},\mathbf{R}_{d})\in\text{SO(3)}\times\text{SO(3)}|\Psi(\mathbf{R},\mathbf{R}_{d})<2 \}
\label{L_2}
\end{IEEEeqnarray}
then, under the assumption of perfect parameter knowledge, we propose the following nonlinear surface-based controller,
\begin{IEEEeqnarray}{rCl}
^{b}\mathbf{u}&=&{}^{b}\boldsymbol{\omega}\times\mathbf{J}{}^{b}\boldsymbol{\omega}-\mathbf{J}\left(\frac{k_{R}}{k_{\omega}}\dot{\mathbf{e}}_{R}+\mathbf{a}_{d}+\eta\mathbf{s}_{R}\right)\label{att_contr}
\end{IEEEeqnarray}
where $\mathbf{a}_{d}$ is defined in App. A(\ref{E_ad}) and the surface $\mathbf{s}_{R}$ is,
\begin{IEEEeqnarray}{C}
\mathbf{s}_{R}=k_{R}\mathbf{e}_{R}+k_{\omega}\mathbf{e}_{\omega}\label{att_surface}
\end{IEEEeqnarray}
Then, the zero equilibrium of the quadrotor closed loop attitude tracking error $(\mathbf{e}_{R},\mathbf{e}_{\omega})=(\mathbf{0},\mathbf{0})$ is almost globally exponentially stable;
moreover there exist constants $\mu,\tau>0$ such that
\begin{IEEEeqnarray}{C}
\Psi(\mathbf{R},\mathbf{R}_{d})<min\{2,\mu e^{-\tau t}\}
\label{Psi_bou}
\end{IEEEeqnarray}

\textbf{Proof.}
See Appendix \ref{appAtt}.

The convergence properties introduced by $\mathbf{s}_{R}$  to the developed attitude controller are analyzed at the end of Section \ref{tracControls} with  the developed position controller.

The initial angular velocity can be arbitrarily large by using sufficiently large gains.
The region of attraction given by (\ref{Psi_0})-(\ref{surface_0}) ensures that the initial attitude error is less than $180^{o}$  with respect to an axis-angle rotation for a desired $\mathbf{R}_{d}$ (i.e., $\mathbf{R}_{d}(t)$ is not antipodal to $\mathbf{R}(t)$).
Consequently exponential stability is guaranteed almost globally.
This is the best that one can do since it has been shown that the topology of SO(3) prohibits the design of a smooth global controller, \cite{obstruction}.

Because (\ref{att_contr}) is developed directly on SO(3), it avoids singularities and ambiguities associated with minimum attitude representations like Euler angles or quaternions completely.
Also this controller can be applied to the attitude dynamics of any rigid body and not only on quadrotor systems.

Since attitude tracking does not depend on $f$, the ACM is more suited for short durations of time.
The thrust magnitude can be selected to achieve an additional objective compatible with the attitude tracking command, i.e. track a desired altitude command \cite{geomquadlee},\cite{geommac},\cite{geomquadlee_asian}.

Finally, despite developing (\ref{att_contr}) under the assumption that $\boldsymbol{\delta}_{R}=0_{3\times1}$, its robustness properties will be tested during simulation in presence of motor saturation and wind disturbances.

\subsection{Position Control Mode (PCM)\label{conmodpos}}
Under the assumption that $\boldsymbol{\delta}_{x}=0_{3\times1}$, a control system is developed for the position dynamics of the quadrotor, stabilizing the tracking errors to zero asymptotically, almost globally.

\subsubsection{Position tracking errors}
For an arbitrary smooth position tracking instruction $\mathbf{x}_{d}\in\mathbb{R}^{3}$, the tracking errors for the position and the velocity are taken as,
\begin{IEEEeqnarray}{C}
\mathbf{e}_{x}=\mathbf{x}-\mathbf{x}_{d},\; \mathbf{e}_{v}=\mathbf{v}-\dot{\mathbf{x}}_{d}\label{pos_error}
\end{IEEEeqnarray}
For $k_{x},k_{v}{\in}\mathbb{R}^{+}$ the position nonlinear surface is defined as,
\begin{IEEEeqnarray}{C}
\mathbf{s}_{x}=k_{x}\mathbf{e}_{x}+k_{v}\mathbf{e}_{v}\label{pos_surf}
\end{IEEEeqnarray}

In the PCM, the attitude dynamics must be compatible with the desired position tracking.
This results in the definition of a position-induced attitude matrix, $\mathbf{R}_{x}(t){\in}\text{SO(3)}$, for use as an attitude command.
To define this matrix, first the desired thrust direction of the  quadrotor, $\mathbf{e}_{3_{x}}$, is computed by,
\begin{IEEEeqnarray}{C}
\mathbf{e}_{3_{x}}{=}\frac{mg\mathbf{E}_{3}-m\frac{k_{x}}{k_{v}}\mathbf{e}_{v}-a\mathbf{s}_{x}+m\ddot{\mathbf{x}}_{d}}{\lVert{mg\mathbf{E}_{3}-m\frac{k_{x}}{k_{v}}\mathbf{e}_{v}-a\mathbf{s}_{x}+m\ddot{\mathbf{x}}_{d}}\rVert}\in\text{S}^{2}, a{\in}\mathbb{R}^{+}\label{e_3_x}
\end{IEEEeqnarray}
where it is assumed that by selecting ${\mathbf{x}}_{d}$, $\dot{\mathbf{x}}_{d}$, $\ddot{\mathbf{x}}_{d}$ hereafter,
\begin{IEEEeqnarray*}{C}
{\lVert{mg\mathbf{E}_{3}-m\frac{k_{x}}{k_{v}}\mathbf{e}_{v}-a\mathbf{s}_{x}+m\ddot{\mathbf{x}}_{d}}\rVert}> 0
\end{IEEEeqnarray*}
Secondly the user defines a desired yaw direction $\mathbf{e}_{1_{d}}\in\text{S}^{2}$ of the $\mathbf{e}_{1}$ body-fixed axis of the quadrotor such that $\mathbf{e}_{1_{d}}\nparallel\mathbf{e}_{3_{x}}$.
This is used to find the position-induced heading, $\mathbf{e}_{1_{h}}$, \cite{geommac}, 
\begin{IEEEeqnarray*}{C}
\mathbf{e}_{1_{h}}=\frac{(\mathbf{e}_{3_{x}}\times\mathbf{e}_{1_{d}})\times\mathbf{e}_{3_{x}}}{\lVert(\mathbf{e}_{3_{x}}\times\mathbf{e}_{1_{d}})\times\mathbf{e}_{3_{x}}\rVert}
\end{IEEEeqnarray*}
The position related attitude $\mathbf{R}_{x}(t){\in}\text{SO(3)}$, $^{b}\boldsymbol{\omega}_{x}(t){\in}\mathbb{R}^{3\times1}$ is,
\begin{IEEEeqnarray}{C}
\mathbf{R}_{x}{=}\left[\mathbf{e}_{1_{h}},\frac{\mathbf{e}_{3_{x}}\times\mathbf{e}_{1_{h}}}{\lVert\mathbf{e}_{3_{x}}\times\mathbf{e}_{1_{h}}\rVert},\mathbf{e}_{3_{x}}\right],\;{}^{b}\boldsymbol{\omega}_{x}{=}{S^{-1}}\!(\mathbf{R}^{T}_{x}\dot{\mathbf{R}}_{x})\label{Rxbox}
\end{IEEEeqnarray}
and the attitude dynamics are guided to follow $\mathbf{R}_{x}(t)$, $^{b}\boldsymbol{\omega}_{x}(t)$.

\subsubsection{Position tracking controller\label{sec:thr_mag}}
Under the assumption that $\boldsymbol{\delta}_{x}=0_{3\times1}$, a control system is developed for the position dynamics of the quadrotor UAV, achieving almost global asymptotic stabilization of ($\mathbf{e}_{x}$,$\mathbf{e}_{v}$,$\mathbf{e}_{R}$,$\mathbf{e}_{\omega}$) to the zero equilibrium through the action/effect of the soon to be introduced Propositions \arabic{Prop4} and \arabic{Prop5}.

For a sufficiently smooth pointing direction $\mathbf{e}_{1_{d}}(t)\in\text{S}^{2}$, and a sufficiently smooth position tracking instruction $\mathbf{x}_{d}(t)\in\mathbb{R}^{3}$ the following position controller is defined,
\begin{IEEEeqnarray}{rCl}
\addtocounter{equation}{1}
\!\!\!\!f(\mathbf{x}_{d},\dot{\mathbf{x}}_{d},\ddot{\mathbf{x}}_{d})&{=}&(mg\mathbf{E}_{3}{-}m\frac{k_{x}}{k_{v}}\mathbf{e}_{v}{-}a\mathbf{s}_{x}{+}m\ddot{\mathbf{x}}_{d})^{T}\mathbf{R}\mathbf{e}_{3}\IEEEyessubnumber\label{f}\\
\!\!\!\!{}^{b}\mathbf{u}(\mathbf{R}_{x},{{}^{b}\boldsymbol{\omega}_{x}})&{=}&{}^{b}\boldsymbol{\omega}{\times}\mathbf{J}{}^{b}\boldsymbol{\omega}{-}\mathbf{J}\left(\frac{k_{R}}{k_{\omega}}\dot{\mathbf{e}}_{R_{x}}{+}\mathbf{a}_{d_{x}}{+}\eta\mathbf{s}_{R_{x}}\right)\IEEEyessubnumber\label{att_contrx}
\end{IEEEeqnarray}
where $\mathbf{s}_{R_{x}}$, $\mathbf{a}_{d_{x}}$, are given by (\ref{att_surface}), App. A(\ref{E_ad}), and $\dot{\mathbf{e}}_{R_{x}}$ is given by App. A(\ref{dot_Att_Error}) if $\{(\ref{error_function}),(\ref{att_error})\}$ are used and is given by App. A(\ref{dot_Att_Error_A}) if $\{(\ref{error_function_A}),(\ref{att_error_A})\}$ are used.
The desired attitude matrix that is used in all the components of (\ref{att_contrx}) is given by (\ref{Rxbox}).

The utilization of nonlinear surfaces resulted to the thrust feedback expression, (\ref{f}), comprised by three gains.
However (\ref{f}) can be scaled to a PD form as in \cite{geomquadlee}.
Since (\ref{f}) is paired with the newly developed attitude controller (\ref{att_contrx}), it forms a new PCM controller of improved closed loop response wrt. \cite{geomquadlee}, see Sect. \ref{simulation}, and its behavior/closed-loop stabilization properties are investigated next.

The closed loop system defined by (\ref{eq:position})-(\ref{attitude_kinem}) under the action of (\ref{f})-(\ref{att_contrx}) is shown to achieve almost global asymptotic stabilization of ($\mathbf{e}_{x}$,$\mathbf{e}_{v}$,$\mathbf{e}_{R}$,$\mathbf{e}_{\omega}$) to the zero equilibrium by the combined action of Propositions \arabic{Prop4} and \arabic{Prop5}.
Specifically (\ref{att_contrx}) drives $\mathbf{R}(t)$ to asymptotically track $\mathbf{R}_{x}(t)$ and combined with (\ref{f}), asymptotic position tracking is achieved.
The first result of exponential stability for a sub-domain of the quadrotor closed loop position dynamics is presented next.

\textbf{Proposition \arabic{Prop4}.}
Considering the controllers in (\ref{f}), (\ref{att_contrx}) and for initial conditions in the domain,
\begin{IEEEeqnarray}{rCl}
D_{x}&=&\{(\mathbf{e}_{x},\mathbf{e}_{v},\mathbf{e}_{R},\mathbf{e}_{\omega})\in\mathbb{R}^{3}\times\mathbb{R}^{3}\times\mathbb{R}^{3}\times\mathbb{R}^{3}|\IEEEnonumber\\
&{ }&\Psi(\mathbf{R}(0),\mathbf{R}_{x}(0))<\psi_{p}<1\}
\label{D_x}
\end{IEEEeqnarray}
and for $\ddot{\mathbf{x}}_{d}\in\mathbb{R}^{3\times1}$, $B\in\mathbb{R}^{+}$ such that the following holds, 
\begin{IEEEeqnarray}{C}
\lVert mg\mathbf{E}_{3}+m\ddot{\mathbf{x}}_{d}\rVert \leq B \label{B}
\end{IEEEeqnarray}
We define $\mathbf{\Pi}_{1},\mathbf{\Pi}_{2}\in\mathbb{R}^{2\times2}$ as,
\begin{IEEEeqnarray}{C}
\mathbf{\Pi}_{1}{=}
\begin{bmatrix}
ak_{x}^{2}(1{-}\theta)&-ak_{x}k_{v}\theta{-}\frac{mk_{x}^{2}\theta}{2k_{v}}\\
-ak_{x}k_{v}\theta{-}\frac{mk_{x}^{2}\theta}{2k_{v}}&ak_{v}^{2}{-}\theta(mk_{x}{+}ak_{v}^{2})
\end{bmatrix},\IEEEnonumber\\
\mathbf{\Pi}_{2}=
\begin{bmatrix}
Bk_{x}&0\\
Bk_{v}&0
\end{bmatrix}\label{P}
\end{IEEEeqnarray}
where $\theta<\theta_{max}\in\mathbb{R}^{+}$ and $\theta_{max}$ is given by,
\begin{IEEEeqnarray}{rCl}
\theta_{max}&=&\min\{\frac{ak_{v}^{2}}{ak_{v}^{2}{+}mk_{x}},\delta_{1}+\delta_{2}\},\IEEEyesnumber\label{theta}\\
\delta_{1}&=&2{\frac {k_{v}^{2}\sqrt {4k_{x}^{4}k_{v}^{4}a^{4}+4k_{x}^{5}k_{v}^{2}{a
}^{3}m+2k_{x}^{6}{m}^{2}{a}^{2}}}{k_{x}^{4}{m}^{2}}}
\IEEEnonumber\\
\delta_{2}&=&-4{\frac {{a}^{2}k_{v}^{4}}{{m}^{2}k_{x}^{2}}}{-}2{\frac {ak_{v}^{2}}{mk_{x}}}\IEEEnonumber
\end{IEEEeqnarray}
If $\{(\ref{error_function}),(\ref{att_error})\}$ is used, the attitude error bound, $\psi_{p}$, satisfies,
\begin{IEEEeqnarray*}{C}
\theta_{max}=\sqrt{\psi_{p}(2-\psi_{p})}
\end{IEEEeqnarray*}
while if the set $\{(\ref{error_function_A}),(\ref{att_error_A})\}$ is used, $\psi_{p}$ satisfies,
\begin{IEEEeqnarray*}{C}
\theta_{max}=\sqrt{\psi_{p}(1-\frac{\psi_{p}}{4})}
\end{IEEEeqnarray*}
In conjunction with suitable gains $\eta,k_{R},k_{\omega}\in\mathbb{R}^{+}$, such that,
\begin{IEEEeqnarray}{C}
\lambda_{min}(\mathbf{W}_{3})>\frac{\lVert\mathbf{\Pi}_{2}\rVert^{2}}{4\eta\lambda_{min}(\mathbf{\Pi}_{1})},\mathbf{W}_{3}=\begin{bmatrix}
k_{R}^{2}&0\\
0&k_{\omega}^{2}
\end{bmatrix}\label{w3}
\end{IEEEeqnarray}
then the zero equilibrium of the closed loop errors $(\mathbf{e}_{x},\mathbf{e}_{v},\mathbf{e}_{R},\mathbf{e}_{\omega})$ is exponentially stable in the domain given by (\ref{D_x}). 
A region of attraction is identified by (\ref{D_x}), (\ref{theta}), and
\begin{IEEEeqnarray}{C}
\lVert\mathbf{e}_{\omega}(0)\rVert^{2}<2\eta k_{R}\left(\psi_{p}-\Psi(\mathbf{R}(0),\mathbf{R}_{x}(0))\right)\label{ep_0}
\end{IEEEeqnarray}
\textbf{Proof.}
See Appendix \ref{appB}.

Proposition \arabic{Prop4} requires that the norm of the initial attitude error is less than $\theta_{max}$ to achieve exponential stability (the upper bound of $\theta$, (\ref{theta}), depends solely on the control gains and the quadrotor mass).
This corresponds to a slightly reduced region of attraction in comparison to the regions in \cite{geomquadlee}, \cite{geommac} -\cite{qeopidfar}, because no restriction on the initial position/velocity error was applied during the stability proof.
This approach is not only novel, wrt. the geometric quadrotor literature, but it also offers the advantage of simplifying the trajectory design procedure.
In contrast, the region of attraction in other geometric treatments includes bounds on the initial position or velocity (see \cite{geomquadlee}, \cite{geommac} -\cite{qeopidfar}) meaning that the trajectory should comply to the position/velocity bounds and also to the attitude bound, a more involved/complicated task.

If a user prefers a larger basin of exponential stability, this can be achieved by introducing bounds on the initial position/velocity (see Appendix \ref{appB}, Section (\ref{altROA}) for more details).
Then two new regions of attraction are produced involving larger initial attitude errors and are given by (\ref{ep_0}) and,
\begin{IEEEeqnarray}{C}
\Psi(\mathbf{R}(0),\mathbf{R}_{x}(0))<\psi_{p}<1, \lVert\mathbf{e}_{x/v}(0)\rVert<e_{x/v_{max}}\label{fra_the_xv}\\
\theta<\theta_{max}=\min\{\frac{ak_{v}^{2}}{ak_{v}^{2}{+}mk_{x}}\}\label{thetaxv}
\end{IEEEeqnarray}
where the second inequality in (\ref{fra_the_xv}) denotes either a bound on the initial position error, $e_{x_{max}}$, or a bound on the initial velocity error, $e_{v_{max}}$, but not on both (see Appendix \ref{appB}, Section (\ref{altROA}) for more details and expressions regarding $\Pi_{1}$, $\Pi_{2}$, that comply with (\ref{w3})).
Depending on user preference, the trajectory design procedure can be realized using either one of the three regions of attraction ($\{(\ref{D_x}), (\ref{theta}), (\ref{ep_0})\}$, $\{(\ref{ep_0}), (\ref{fra_the_xv}), (\ref{thetaxv})\}$ using $e_{x_{max}}$ and $\{(\ref{ep_0}), (\ref{fra_the_xv}), (\ref{thetaxv})\}$ using $e_{v_{max}}$) guiding us to favorable conditions for switching between flight modes.
For completeness, all three regions of exponential stability were derived;
however this work focuses on the region given by $\{(\ref{D_x}), (\ref{theta}), (\ref{ep_0})\}$.

Finally, the proposition that follows shows that the structure of the position controller is characterized by almost global exponential attractiveness.
This compensates for the reduced position/velocity free region of attraction and introduces greater freedom to the user in regards to control objectives, since the region of attraction does not depend explicitly on the initial position/velocity error.
If the quadrotor initial states are outside of (\ref{D_x}), with respect to the initial attitude, Proposition \arabic{Prop3} still applies due to the action of (\ref{att_contrx}).
Thus the attitude state enters (\ref{D_x}) in finite time $t^{*}$ and the results of Proposition \arabic{Prop4} take effect.
The result regarding the position mode is stated next.

\textbf{Proposition \arabic{Prop5}.}
For initial conditions satisfying (\ref{surface_0}), and
\begin{IEEEeqnarray}{C}
\psi_{p}\leq\Psi(\mathbf{R}(0),\mathbf{R}_{x}(0))<2\label{Psi_atr}
\end{IEEEeqnarray}
and a uniformly bounded desired acceleration (\ref{B}), the thrust magnitude defined in (\ref{f}), in conjunction with the control moment (\ref{att_contrx}), renders the zero equilibrium of $(\mathbf{e}_{x},\mathbf{e}_{v},\mathbf{e}_{R},\mathbf{e}_{\omega})$ almost globally exponentially attractive.

\textbf{Proof of Proposition \arabic{Prop5}.}
See Proposition 4 in \cite{geommac} but apply the thrust feedback expression (\ref{f}).

Proposition \arabic{Prop5} shows that during the finite time that it takes for the attitude states to enter the region of attraction for exponential stability (\ref{D_x}), (\ref{ep_0}), the position tracking errors (\ref{pos_error}) remain bounded.
The calculated region of exponential attractiveness given by (\ref{Psi_atr}) ensures that the initial attitude error is less than $180^{o}$  with respect to an axis-angle rotation for a desired $\mathbf{R}_{x}$ (i.e., $\mathbf{R}_{x}(t)$ is not antipodal to $\mathbf{R}(t)$).
Consequently the zero equilibrium of the tracking errors is almost globally exponentially attractive.

\begin{figure}[!h]
\centering
\subfloat[\label{attsur}]{\includegraphics[width=0.49\columnwidth]{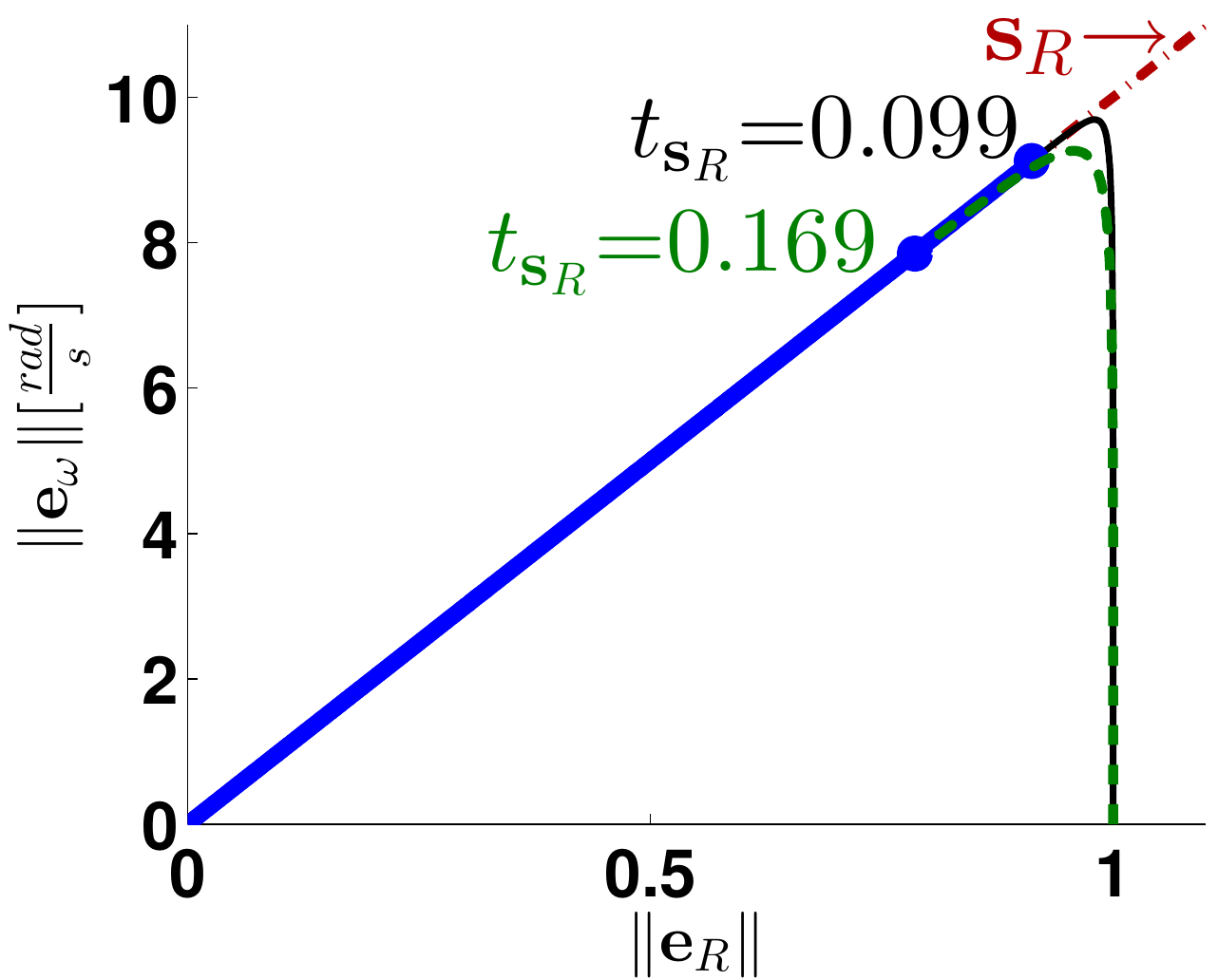}}
~
\subfloat[\label{possur}]{\includegraphics[width=0.49\columnwidth]{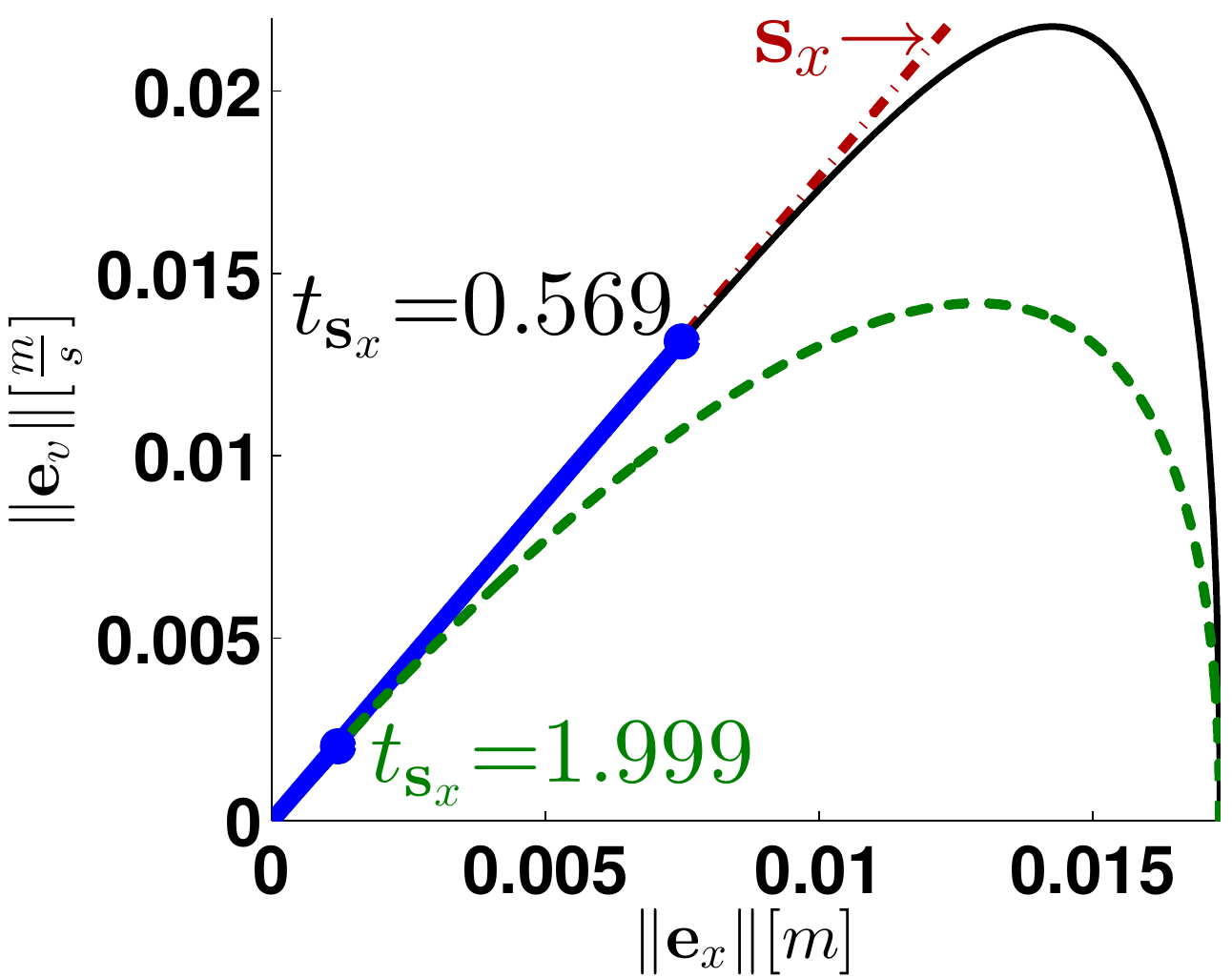}}

\caption{
Sliding behavior produced by, (\ref{att_contr}), ((\ref{f}), (\ref{att_contrx})) using $\{(\ref{error_function_A}),(\ref{att_error_A})\}$.
(\ref{attsur}) Convergence to $\mathbf{s}_{R}$ for a step of $179.9999^{o}$.
(\ref{possur}) Convergence to $\mathbf{s}_{x}$ for a position step to $\mathbf{x}_{d}{=}[1;1;1]cm$.
The black and dashed green lines indicate the reaching phase to $\mathbf{s}_{R,x}$ followed by sliding behavior indicated by blue lines.
The black lines indicate usage of higher sliding gains $\eta,a$.
The reaching times, $t_{\mathbf{s}_{R,x}}$, are colored accordingly.
}
\label{SMCc}
\end{figure}

Note that for both control modes \ref{conmodatt} (\ref{conmodpos}), through the utilization of the nonlinear surfaces $\mathbf{s}_{R}$, ($\mathbf{s}_{x}$), the closed loop dynamics of the nonlinear system are altered, enabling the user to influence the convergence of the system to the zero equilibrium by using three gains per surface.
First by using the gains $\eta$, ($a$), to affect the reaching time to the surface, by penalizing the combined surface error, followed by the gains $k_{R},k_{\omega},(k_{x},k_{v})$, to affect the convergence time when on/near the surface by penalizing independently the attitude, angular velocity, (position, translational velocity), errors.
This is showcased in  Fig. \ref{SMCc}, where the quadrotor response is shown during an attitude maneuver (Fig. \ref{attsur}), and a position maneuver (Fig. \ref{possur}).
In both cases, the same simulation is repeated but with larger gains $\eta$, ($a$), resulting in faster reaching times, see black solid lines in Fig. \ref{attsur},\ref{possur}.
In Fig. \ref{attsur}, by doubling $\eta$, the reaching time	 from $t_{\mathbf{s}_{R}}{=}0.169$ improves to $t_{\mathbf{s}_{R}}{=}0.099$ and in Fig. \ref{possur}, by increasing $a$ by four, the reaching time from $t_{\mathbf{s}_{x}}{=}1.999$ improves to $t_{\mathbf{s}_{x}}{=}0.569$.
As a result, the strict algebraic relation to the gains imposed by the proposed controller design, introduces ''sliding like'' closed loop dynamics, see description in Fig. \ref{SMCc}, and allows for finer control on the convergence rate to the zero equilibrium by using the insights gained by the Lyapunov analysis.
Also the sliding behavior is achieved here without the signum function; thus chattering is avoided.

Due to the combined action of (\ref{f}) with (\ref{att_contrx}) it was possible to identify, for the first time wrt. the geometric literature, a region of attraction \textit{independent} of the initial position/velocity error.
This is a new development in regards to the geometric literature.
Additionally the developed expression, (\ref{f}), with the third gain allows for more intuitive tuning thus offering further refinement of the closed loop response.

Concluding, by the combined action of Propositions \arabic{Prop4} and \arabic{Prop5}, asymptotic almost global stabilization of ($\mathbf{e}_{x}$,$\mathbf{e}_{v}$,$\mathbf{e}_{R}$,$\mathbf{e}_{\omega}$) to the zero equilibrium is achieved.
Since both flight modes have almost global stability properties, the closed loop system is robust to switching between flight modes.
The only consideration in respect to trajectory planning is that the desired trajectory must agree with (\ref{Psi_0})-(\ref{surface_0}).
Despite developing (\ref{f}), (\ref{att_contrx}) under the assumption that $\boldsymbol{\delta}_{x}{=}0_{3\times1}$, the robustness of the controller will be tested during simulation in the presence of motor saturation and wind disturbances.

\section{Results\label{simulation}}
The effectiveness of the developed GNCS is verified through simulations.
First by a comparison with the GNCS in \cite{geomquadlee}, to verify the claims from Section \ref{sec:thr_mag} in regards to the thrust magnitude (\ref{f}), followed by an aggressive recovery/trajectory tracking maneuver in the presence of motor saturations and noise to test the effectiveness and robustness of the developed GNCS.

To analyze GNCSs consisting of different structure and strategies, a criterion is needed for a commensurate comparison of their performance.
To this end the Root-Mean-Square (RMS) of the thrusts is used as a criterion, given by,
\begin{IEEEeqnarray}{C}
f_{RMS}(t)=\sqrt{\frac{1}{t}\int_{0}^{t}\sum_{1}^{4} [f_{i}(t)]^{2} d\tau}
\label{rms}
\end{IEEEeqnarray}
Specifically we use (\ref{rms}) to calculate the RMS control effort difference, ${\Delta}f_{RMS}(t)$, given by,
\begin{IEEEeqnarray}{C}
{\Delta}f_{RMS}(t)=f^{proposed}_{RMS}(t)-f^{benchmark}_{RMS}(t)
\label{Deltarms}
\end{IEEEeqnarray}
and tune our developed GNCS such that (\ref{Deltarms}) is negative during the simulation at all times so that the benchmark controller has equal or larger control authority.
By comparing the controller performance, if the developed GNCS produces the least error with \textit{less control effort} it is deemed superior.
The system parameters were taken from a real quadrotor described in \cite{Hinf}:
\begin{IEEEeqnarray*}{C}
\mathbf{J}=[0.0181,0,0;0,0.0196,0;0,0,0.0273]\;kg\,m^{2}\IEEEnonumber\\
m=1.225\;kg,
d=0.23\;m,
b_{T}=0.0121\;m\IEEEnonumber
\end{IEEEeqnarray*}
and the actuator constraints, see \cite{Hinf}, are given by:
\begin{IEEEeqnarray*}{C}
f_{i,min}=0{ }\text{[N]}, f_{i,max}=6.9939{ }\text{[N]}
\end{IEEEeqnarray*}
The wind profile shown in Fig. \ref{windpr} is used in conjunction with the drag equation, \cite{Batchelor}, with the drag coefficient and reference area matrices of the quadrotor to be given by,
\begin{IEEEeqnarray*}{C}
C_{D}{=}\text{diag}(0.2{,}0.22{,}0.5),
A_{D}{=}\text{diag}(0.0907{,}0.0907{,}0.4004)\text{m}^2
\end{IEEEeqnarray*}
The torque due to wind is calculated by assuming that the disturbance force is applied at $0.04\mathbf{e}_{3}$.
Finally all simulations were conducted using fixed-step integration with $dt{=}1{\cdot}10^{-3}$s.

\subsection{Geometric-NCS comparison}
\begin{figure}[!h]
\centering
\subfloat[\label{frmstun}]{\includegraphics[width=0.5\columnwidth]{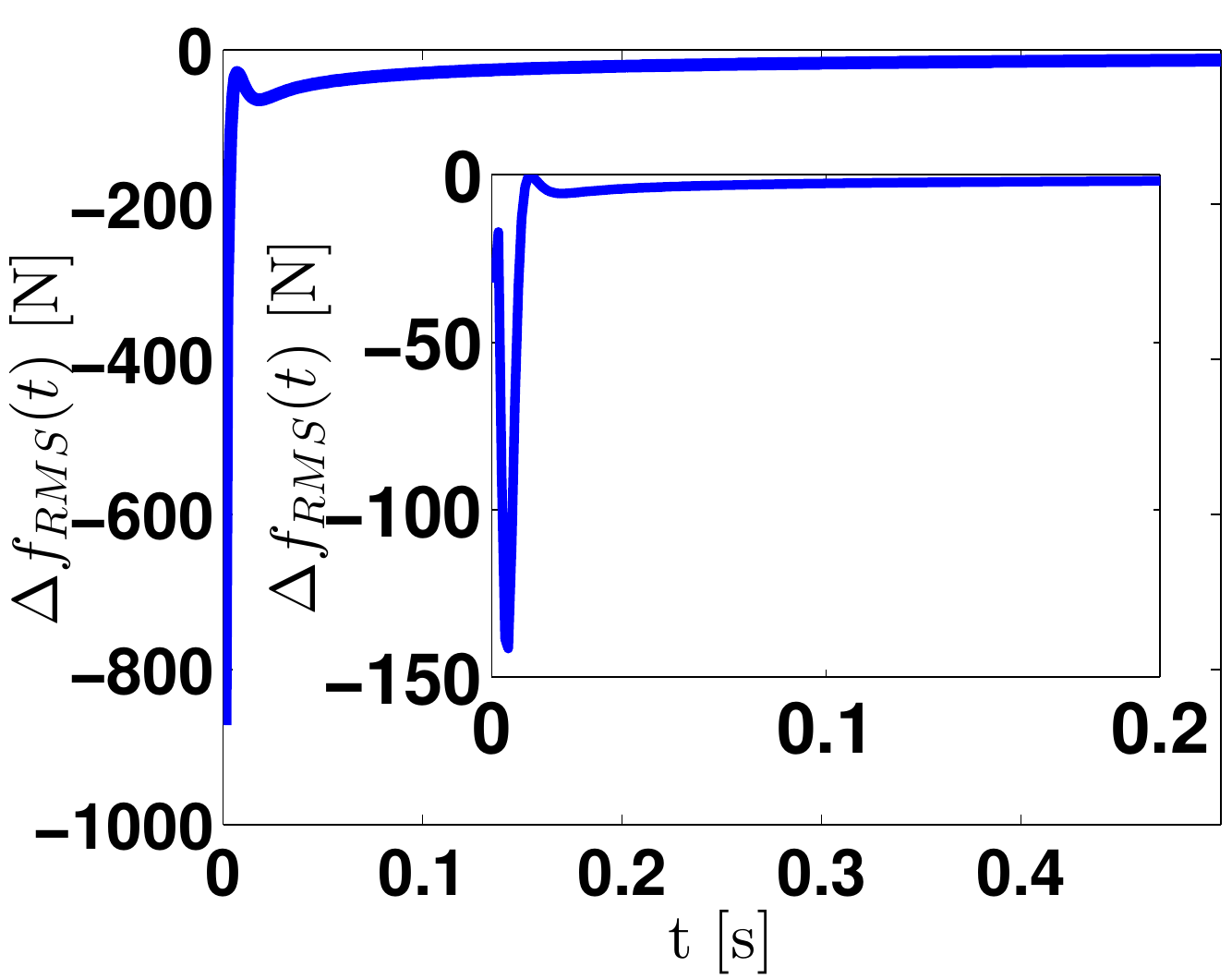}}
~
\subfloat[\label{psi90tun}]{\includegraphics[width=0.5\columnwidth]{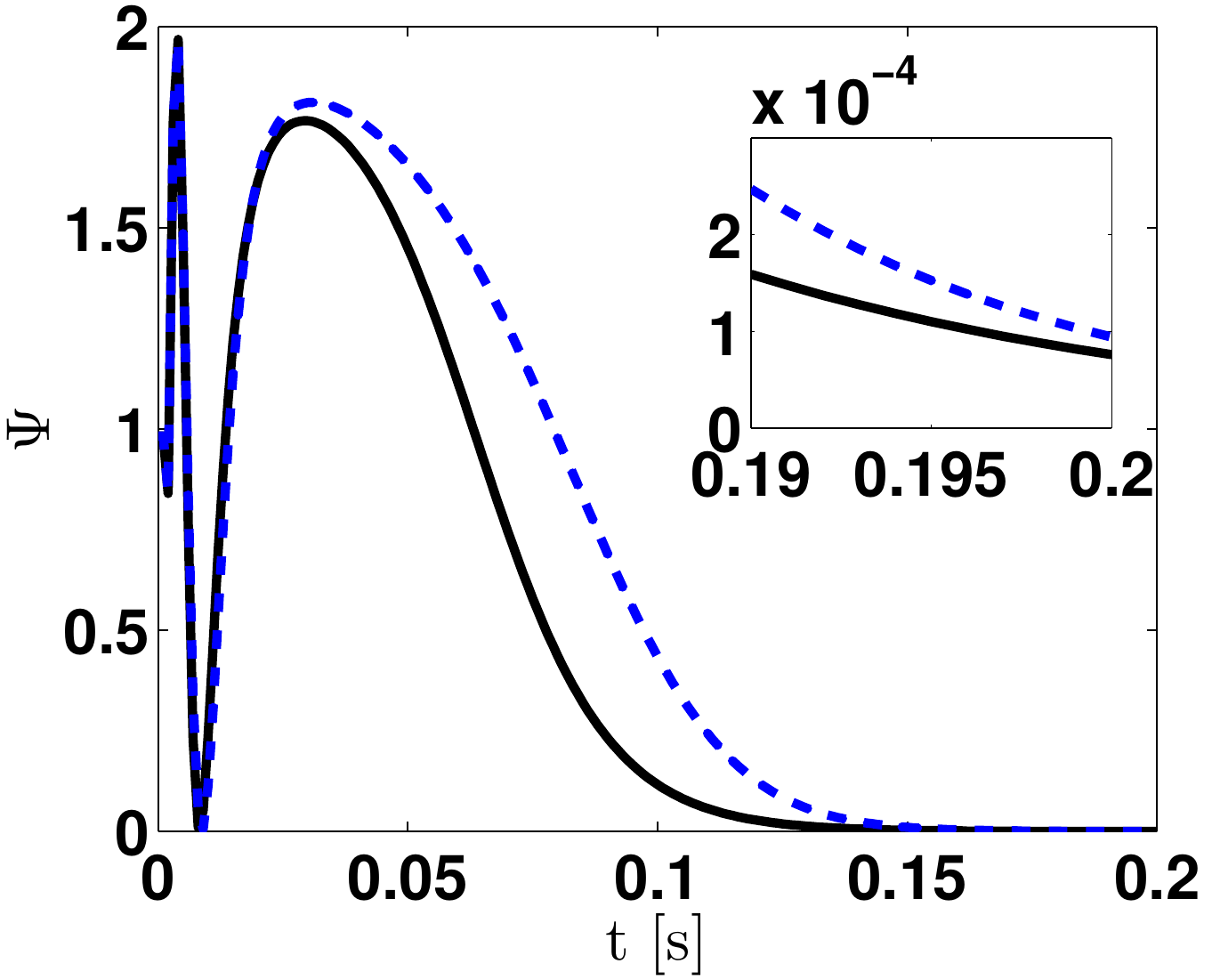}}

\subfloat[\label{psitun}]{\includegraphics[width=0.5\columnwidth]{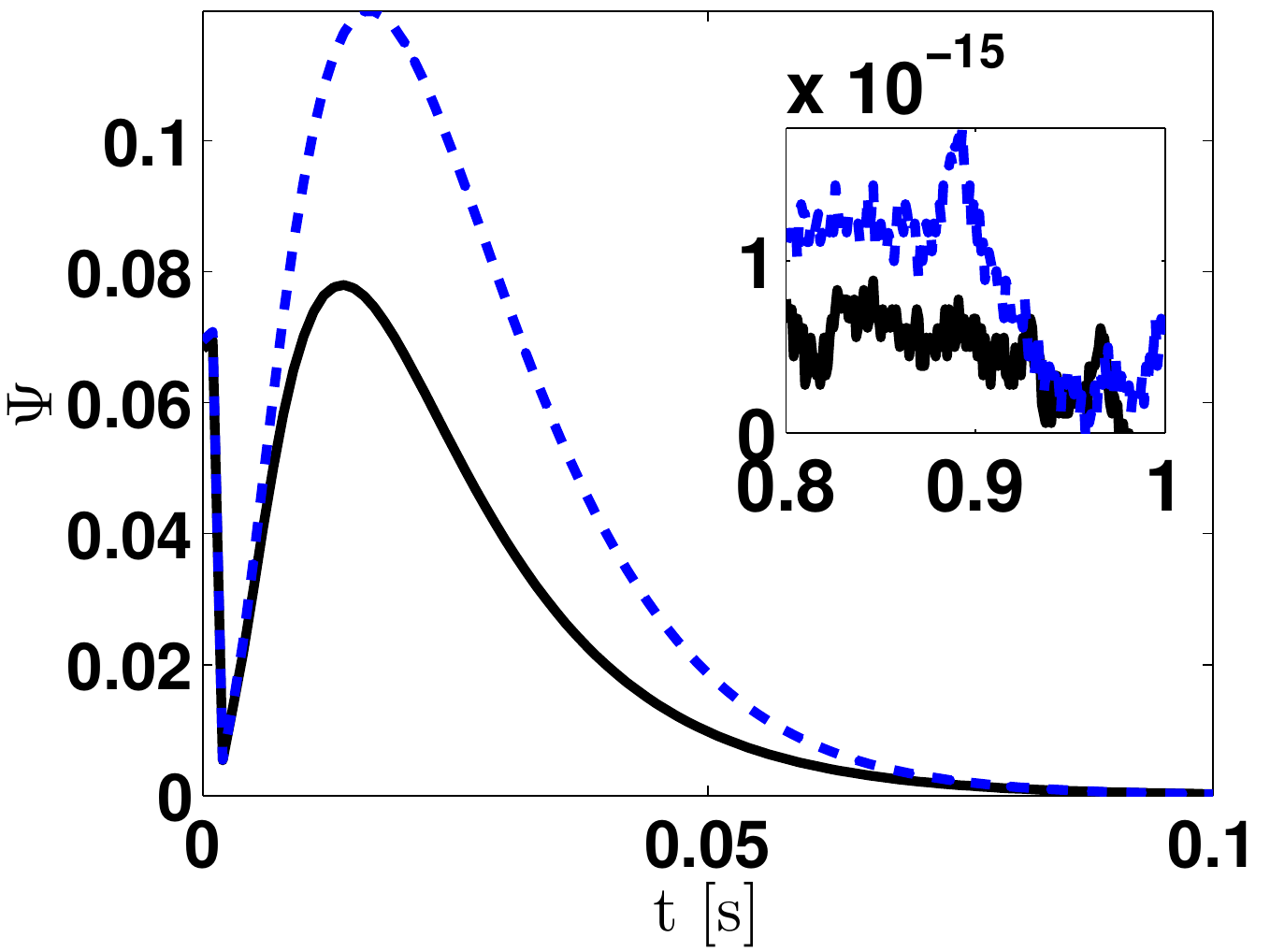}}
~
\subfloat[\label{trajtun}]{\includegraphics[width=0.5\columnwidth]{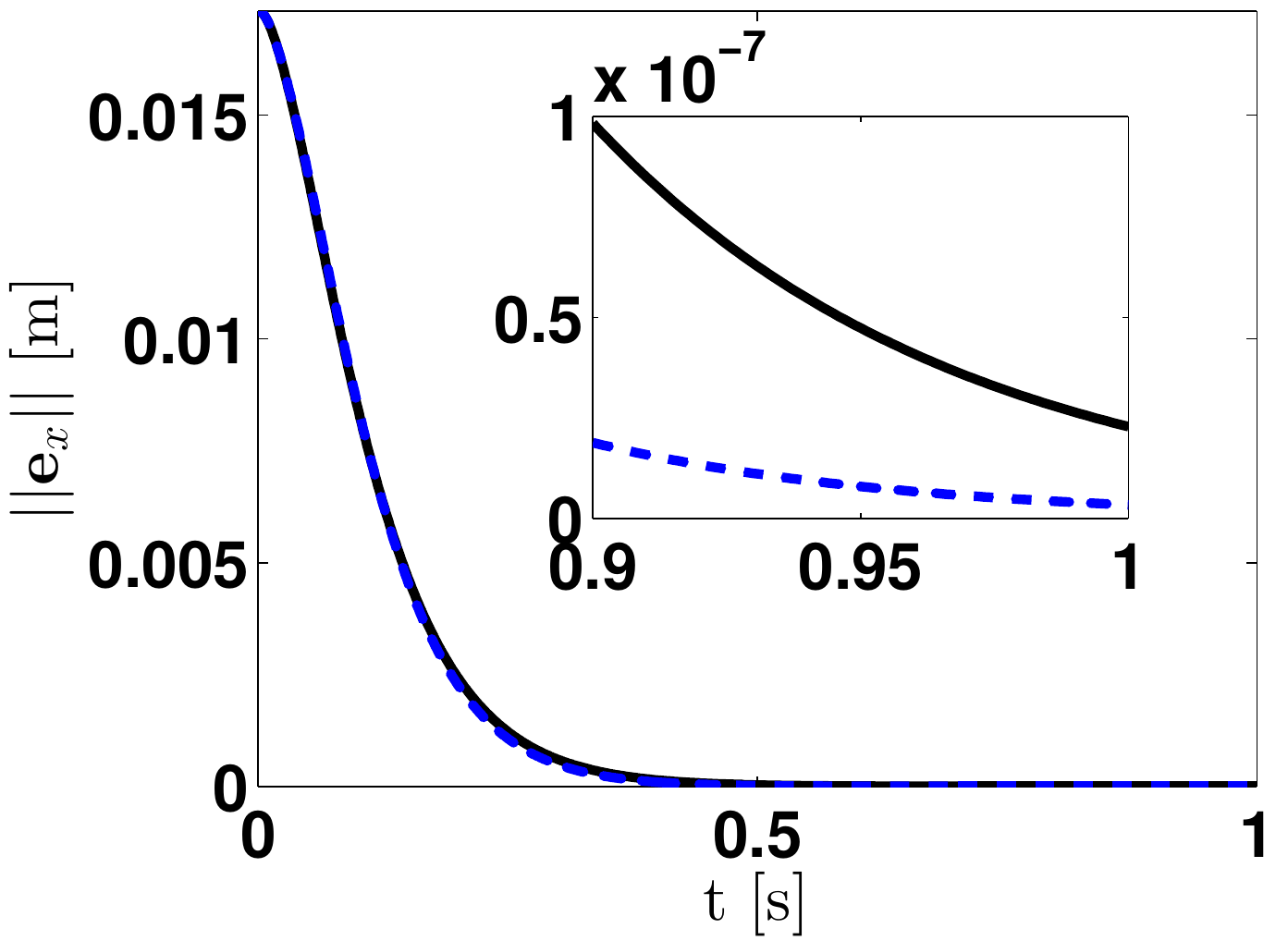}}

\caption{Quadrotor response after the tuning procedure.
(\ref{frmstun}) RMS control effort by (\ref{Deltarms}).
(\ref{psi90tun}) Response for a step command of $90^{o}$.
(\ref{psitun},\ref{trajtun}) Response for a position command to $\mathbf{x}_{d}{=}[1;1;1][cm]$.
(\ref{psitun}) Attitude error given by (\ref{error_function}).
(\ref{trajtun})
Position error, $\lVert\mathbf{e}_{x}\rVert$.
Solid lines (1): Developed, Dashed lines (2): Benchmark.
}
\label{Tuning}
\end{figure}

For this comparison, the GNCS in \cite{geomquadlee} was selected as a benchmark since it is the first quadrotor GNCSs developed directly on SE(3), it demonstrates remarkable results in aggressive maneuvers, and to validate the claims of Sect. \ref{sec:thr_mag}.
The controllers use the first set of error vectors given by $\{(\ref{error_function}),(\ref{att_error})\}$, and no saturation/disturbances are included, to conclude controller competence.
The gains were tuned using (\ref{Deltarms}) as follows.
First the attitude gains were tuned for a desired pitch command of $90^{o}$ followed by tuning the position gains for a desired $\mathbf{x}_{d}{=}[1;1;1][cm]$.
Tuning the attitude controller first, ensures that during the PCM, the attitude controller embedded in the position control loop will produce homogeneous control effort.
Also the gains must be compliant to (\ref{eq:hta}), (\ref{w3}).
The developed controller gains are: 
\begin{IEEEeqnarray*}{C}
k_{\omega}{=}150,
k_{R}{=}5625,
\eta{=}0.8\\
k_{v}{=}59.82,
k_{x}{=}894.62,
a{=}0.5071
\end{IEEEeqnarray*}
The benchmark controller \cite{geomquadlee} parameters used are: 
\begin{IEEEeqnarray*}{C}
k_{\omega}=[2.1720,0,0;0,2.3520,0;0,0,3.2760]\\
k_{R}{=}[65.16,0,0;0,70.56,0;0,0,98.28],k_{v}{=}38.71,k_{x}{=}375.61
\end{IEEEeqnarray*}
The initial conditions (IC's) are: $\mathbf{x}(0)=\mathbf{v}(0)={}^{b}\boldsymbol{\omega}(0)=0_{3\times 1},\mathbf{R}(0)=\mathbf{I}$.
The results  are presented in Fig. \ref{Tuning}.

Examining Fig. \ref{psi90tun}, the effectiveness of  (\ref{att_contr}) (solid black line: 1) with respect to the benchmark controller (dashed blue line: 2) in performing attitude maneuvers is demonstrated as $\Psi$ converges to zero faster and with less control effort, see Fig. \ref{frmstun} inner plot.
The quadrotor response for a position command to $\mathbf{x}_{d}{=}[1;1;1][cm]$ is shown in Fig (\ref{psitun},\ref{trajtun}).
Examining  Fig. (\ref{trajtun}), it is clear that the developed position controller ((\ref{f}), (\ref{att_contrx})) performs equally well with the benchmark controller.
However the attitude error during the position maneuver is negotiated better by the developed position controller as $\Psi$ converges to zero faster and with a smaller overall error, $\Psi{<}0.078$, vs $\Psi{<}0.1198$, an important prevalence.
In Fig. \ref{frmstun} the value of, (\ref{Deltarms}), is displayed for both the attitude (inner plot), and position (outer plot), maneuvers.
Notice that the benchmark controller underperforms despite using more control effort, see Fig. \ref{frmstun}.

The reason that (\ref{Deltarms}) exhibits large values in Fig. \ref{frmstun}, is due to the high gains used to achieve precise trajectory tracking.
As a result because the controllers are fed with step commands, extremely large control efforts are observed.

In view of the above, the ability of the developed PCM in achieving the position command coequally to \cite{geomquadlee} but with less control effort while simultaneously negotiate the attitude error more efficiently again with less control effort makes it more effective and validates the claims of Sect. \ref{sec:thr_mag}.

\subsection{Aggressive recovery/trajectory tracking maneuver}
A complex flight maneuver is conducted, in the presence of motor saturation and noise due to wind, involving transitions between flight modes.
In this simulation, the developed controllers utilize the second set of error vectors given by, $\{(\ref{error_function_A}),(\ref{att_error_A})\}$.
The maneuver was selected to showcase both the trajectory tracking for position and attitude, and the recovery capabilities of the developed GNCS.
The IC's are: $\mathbf{x}(0)=[0;0;5],\mathbf{v}(0)={}^{b}\boldsymbol{\omega}(0)=0_{3\times 1},\mathbf{R}(0)=\mathbf{I}$.
Since this simulation contains portions characterized by large error vectors, softer gains are needed to ensure smooth behavior and minimize motor saturation.
The gains used are: 
\begin{IEEEeqnarray*}{C}
k_{\omega}{=}40,
k_{R}{=}400,
\eta{=}1.002\\
k_{v}{=}7.06,
k_{x}{=}12.46,
a{=}0.5081
\end{IEEEeqnarray*}
The flight scenario, to be achieved through the concatenation of the two flight modes, is described next:
\begin{enumerate}[(a)]
\item ($t < 4$): Position Mode: Translation from the IC's to $\mathbf{x}_{d}=[0;1;10],\mathbf{v}_{d}=[0;0;7],\mathbf{e}_{1d}=[1;0;0]$ using smooth polynomials of eighth degree (SP$8^{th}$).
\item ($4\leq t < 4.4$): Attitude Mode: The quadrotor performs a $180^{o}$ pitch maneuver, i.e. goes inverted.
$\mathbf{R}_{d}(t)$ was designed by defining the pitch angle using SP$8^{th}$.
\item ($4.4\leq t < 4.9$): Attitude Mode: The quadrotor recovers from its inverted state to $\mathbf{R}_{d}(t)=\mathbf{I}$, i.e. point to point command. 
\item ($4.9\leq t \leq 10$): Position Mode: Translation to $\mathbf{x}_{d}=[-1;1.5;10],\mathbf{e}_{1d}=[1;0;0]$ using SP$8^{th}$ with IC's equal to the values of the states of the quadrotor at the end of the attitude mode.
\end{enumerate}

\begin{figure}[!h]
\centering
\subfloat[\label{PsiComp}]{\includegraphics[width=0.5\columnwidth]{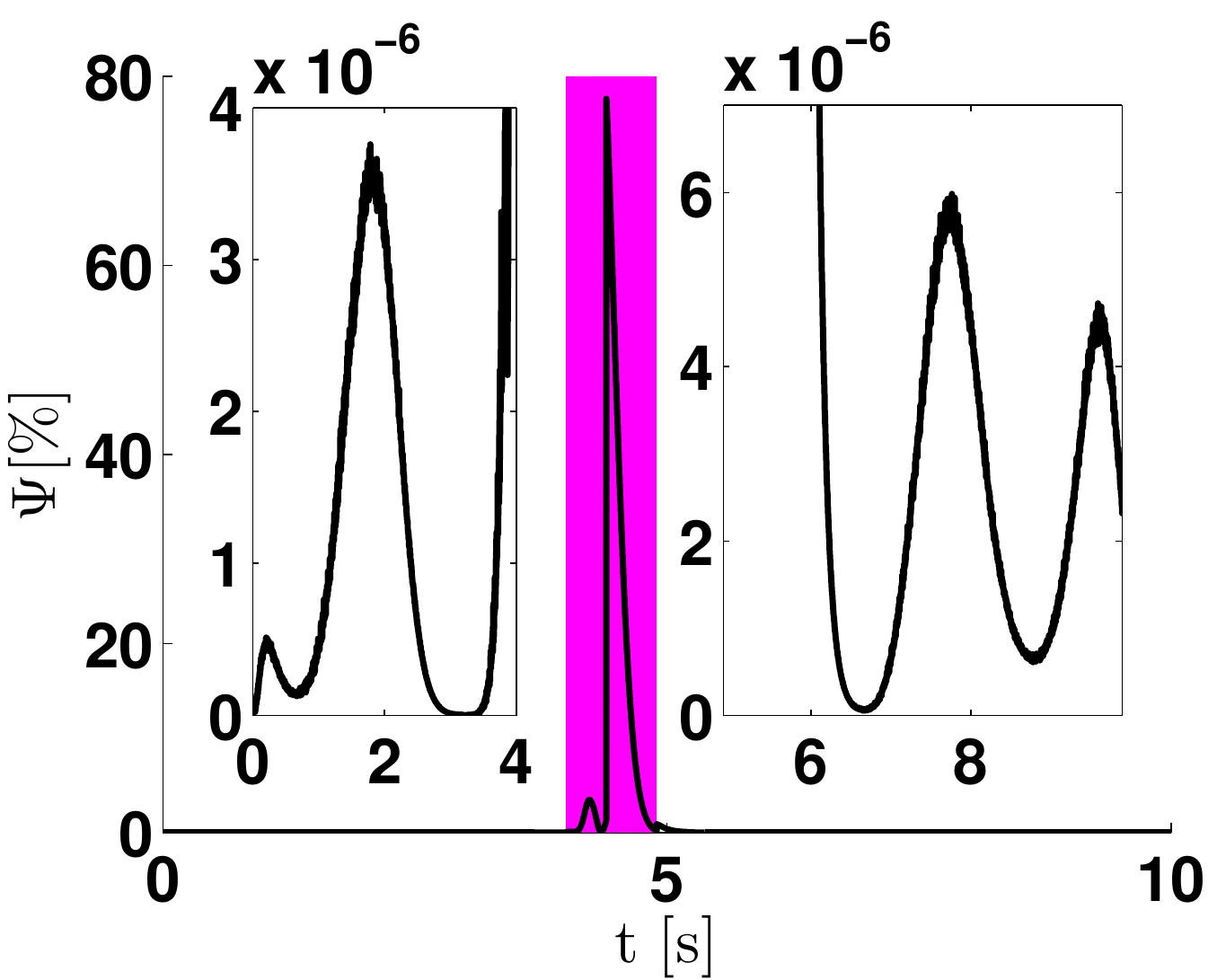}}
~
\subfloat[\label{PosComp}]{\includegraphics[width=0.5\columnwidth]{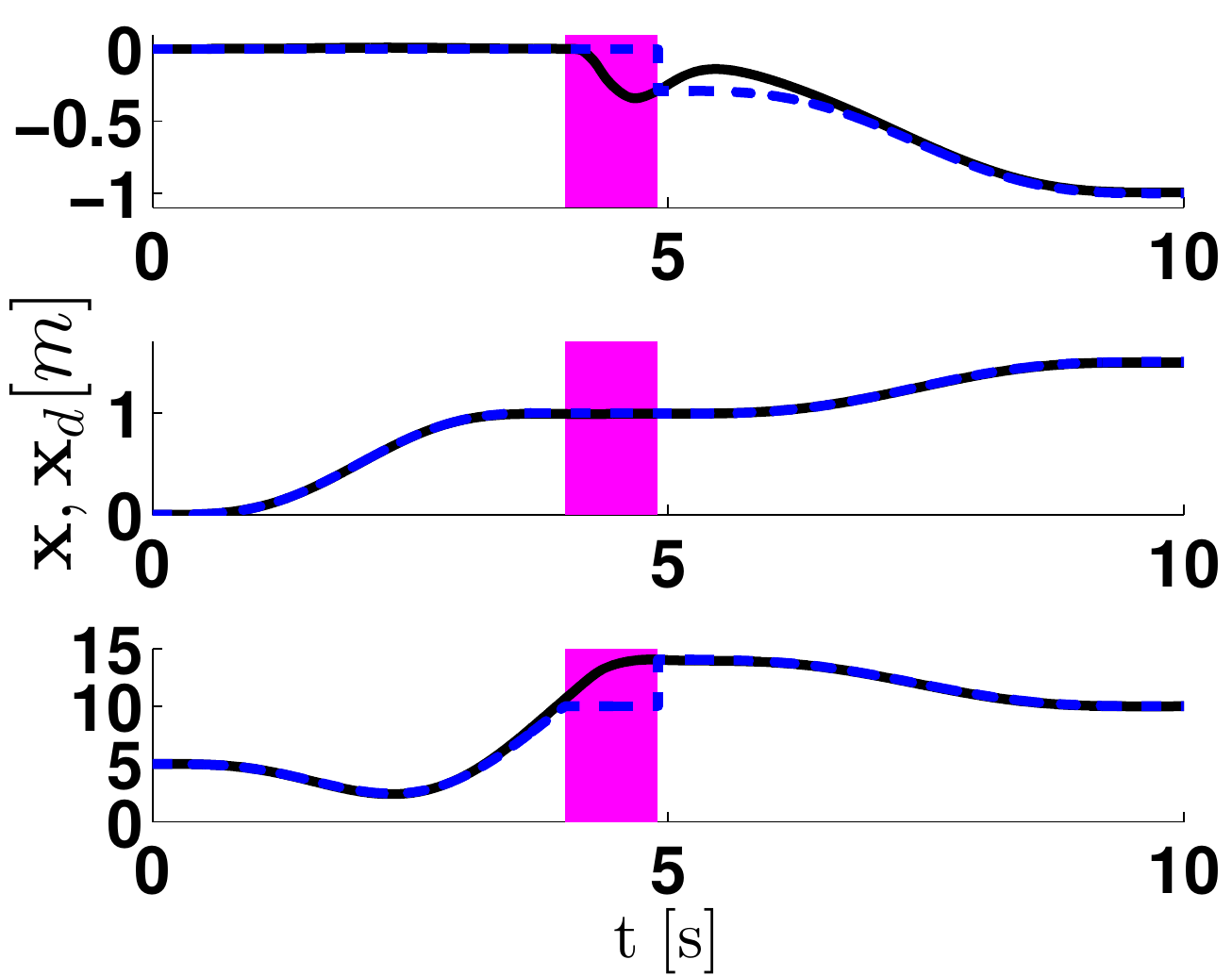}}

\subfloat[\label{thr_U}]{\includegraphics[width=0.5\columnwidth]{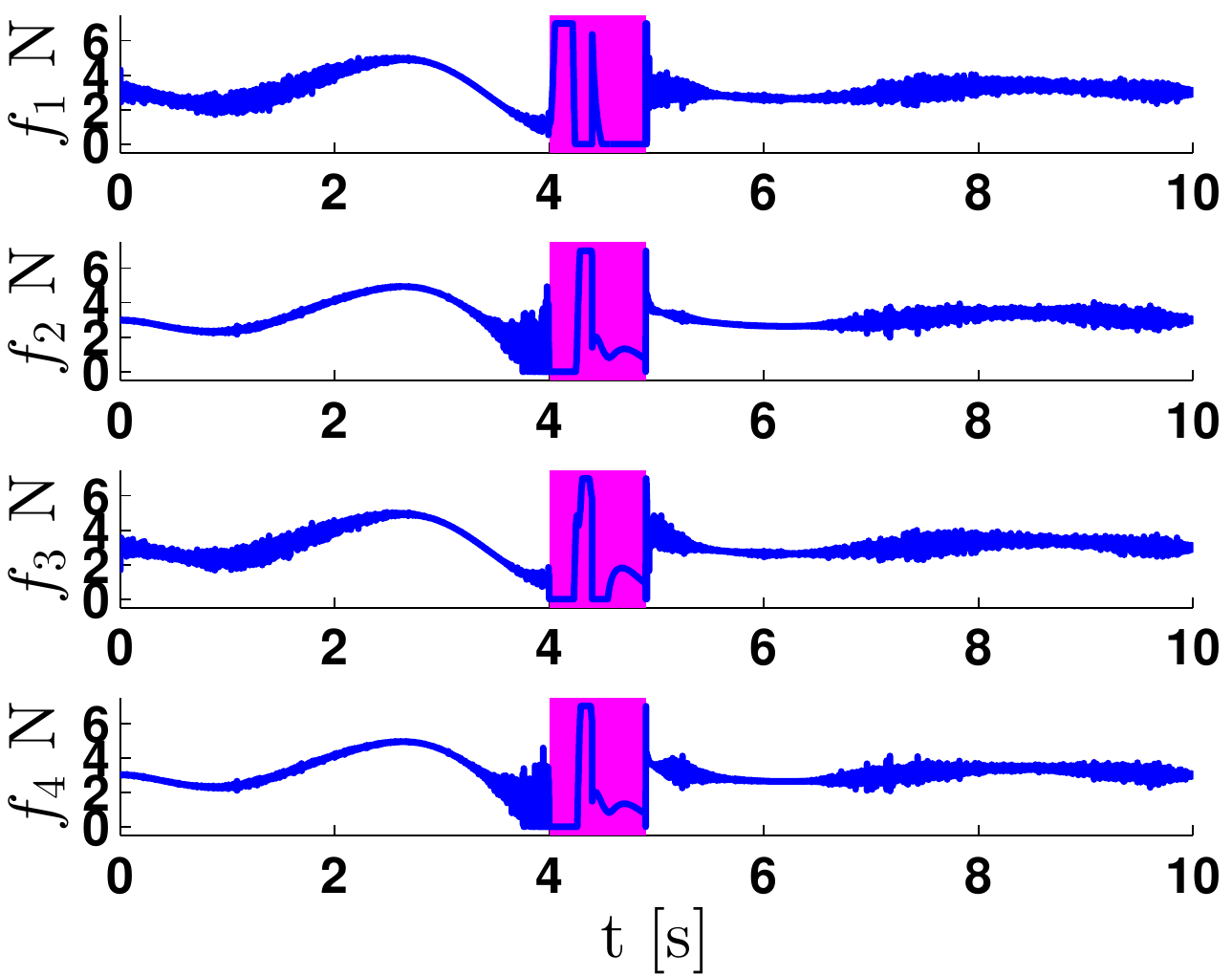}}
~
\subfloat[\label{windpr}]{\includegraphics[width=0.5\columnwidth]{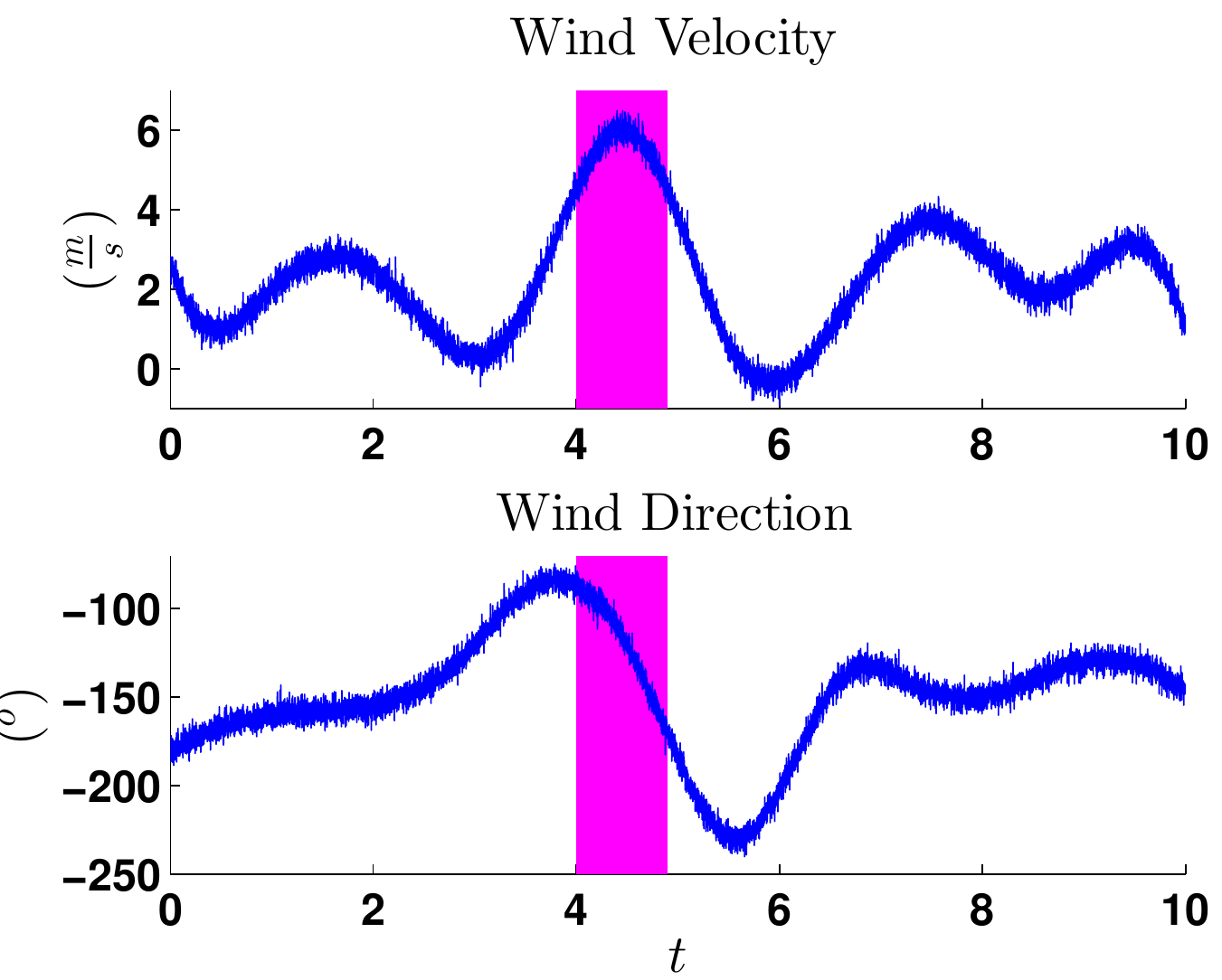}}

\caption{
Complex trajectory tracking.
(\ref{PsiComp}) Attitude error given by (\ref{error_function_A}). 
(\ref{PosComp}) Position state $\mathbf{x}(t)$ (solid black line) and reference $\mathbf{x}_{d}(t)$ (blue dashed line).
(\ref{thr_U}) Thrusts (Developed).
(\ref{windpr}) Wind profile.}
\label{Aggressive}
\end{figure}

Simulation results of the maneuver are illustrated in Fig. \ref{Aggressive} where the duration that the attitude mode is utilized is illustrated by the magenta colored intervals.
The percentage attitude error using (\ref{error_function_A}) is shown in Fig. \ref{PsiComp}.
It is observed that up to $t=4.4$, i.e. the beginning of the quadrotor recovery from the inverted position, the quadrotor atttitude error is maintained below 5\% (below $9^{o}$ wrt. an axis-angle rotation).
During the recovery interval ($4.4<t<4.9$), despite the large attitude error of $77.64\%$ introduced by the attitude step command, the quadrotor successfully converges to the desired orientation undeterred by the disturbances due to wind and motor saturations, see Fig. \ref{thr_U}, \ref{windpr}.
The position response is shown in Fig. \ref{PosComp}.
During the position mode, i.e. $t<4$ and $t>4.9$, the states track the reference trajectories effectively, see Fig. \ref{PosComp}. 
At the position mode interval, $\lVert\mathbf{e}_{x}\rVert$ (not shown here due to space) increases above 0.06m, to 0.5m, only between $3<t<4$ where the wind increases rapidly, see Fig. \ref{windpr} for the wind profile.
The effect of the wind at the same interval is evident also by the noisy motor thrusts, see Fig. \ref{thr_U} at $3<t<4$.
A simulation conducted in the absence of wind, not shown due to space, showed that the noisy behavior in Fig. \ref{thr_U} is eradicated and $\lVert\mathbf{e}_{x}\rVert<0.06$ throughout the position mode interval.
Concluding, the effectiveness of the proposed GNCSs in performing precise trajectory tracking maneuvers (attitude/position) and recovery maneuvers in the presence of motor saturations and disturbances was shown.
The safe switching between flight modes, stated at the end of Section \ref{conmodpos}, was also demonstrated.

\section{Conclusion and Future Work}
In this paper, new controllers for a quadrotor unmanned micro aerial vehicle were developed, based on nonlinear surfaces and employing tracking errors that evolve directly on the nonlinear configuration manifold, inherently including in the control design the nonlinear characteristics of the SE(3) configuration space.
Through rigorous stability proofs, the developed controllers were shown to have desirable closed loop properties that are almost global.
A region of attraction, independent of the position error, was produced and analyzed for the first time, wrt. the geometric literature.
The effectiveness of the developed GNCS was validated by numerical simulations of aggressive maneuvers, in the presence of motor saturations and disturbances due to wind.

Our future work will include experimental trials and an investigation of the developed GNCS robustness properties.


%


%

%
\appendices
\section{\label{appA}}
The attitude tracking errors associated with the attitude error functions studied in \cite{qeoadapclee}, \cite{err_fun}, and related   properties are summarized next.

\textbf{Proposition \arabic{Prop1}.}
Employing $\{(\ref{error_function}), (\ref{att_error})\}$, for a given tracking command $\mathbf{R}_{d}$ and current attitude $\mathbf{R}$, the following hold:
\begin{enumerate}[(i)]
\item $\Psi$ is locally positive-definite about $\mathbf{R}=\mathbf{R}_{d}$ and,
\begin{IEEEeqnarray}{rCl}
\lVert\mathbf{e}_{R}(\mathbf{R},\mathbf{R}_{d})\rVert^{2}&=&(2-\Psi(\mathbf{R},\mathbf{R}_{d}))\Psi(\mathbf{R},\mathbf{R}_{d})
\label{err_eq_psi}
\end{IEEEeqnarray}
\item A lower bound of $\Psi$ is given as follows,
\begin{IEEEeqnarray}{rCl}
\frac{1}{2}\lVert\mathbf{e}_{R}(\mathbf{R},\mathbf{R}_{d})\rVert^{2}&\leq &\Psi(\mathbf{R},\mathbf{R}_{d})
\label{low_Psi}
\end{IEEEeqnarray}
\item Let $\psi\in\mathbb{R}^{+}$.
If $\Psi(\mathbf{R},\mathbf{R}_{d})<\psi<2$, then the upper bound of $\Psi$ is given by,
\begin{IEEEeqnarray}{rCl}
\Psi(\mathbf{R},\mathbf{R}_{d})&\leq &\frac{1}{2-\psi}\lVert\mathbf{e}_{R}(\mathbf{R},\mathbf{R}_{d})\rVert^{2}
\label{upper_Psi}
\end{IEEEeqnarray}
\item The left-trivialized derivative of $\Psi$ is given by,
\begin{IEEEeqnarray}{rCl}
\text{T}^{*}_{I}\text{L}_{R}(\mathbf{D}_{R}\Psi(\mathbf{R},\mathbf{R}_{d}))=\mathbf{e}_{R}
\label{left_triv_der}
\end{IEEEeqnarray}
\item The critical points of $\Psi$, where $\mathbf{e}_{R}=0$, are $\{\mathbf{R}_{d}\}\cap\{\mathbf{R}_{d}\text{exp}(\pi S(\mathbf{s})),\mathbf{s}\in\text{S}^{2}\}$.
\setcounter{numberlatin}{\value{enumi}}
\end{enumerate}

As to $\{(\ref{error_function_A}), (\ref{att_error_A})\}$, the attitude error vector is well defined in (\ref{L_2}).
Thus for a tracking command $\mathbf{R}_{d}$ and current attitude $\mathbf{R}$,
\begin{enumerate}[(i)]
\setcounter{enumi}{\value{numberlatin}}
\item $\Psi$ is locally positive-definite about $\mathbf{R}=\mathbf{R}_{d}$.
\item In (\ref{L_2}) the left-trivialized derivative of $\Psi$ is given by,
\begin{IEEEeqnarray}{rCl}
\text{T}^{*}_{I}\text{L}_{R}(\mathbf{D}_{R}\Psi(\mathbf{R},\mathbf{R}_{d}))=\mathbf{e}_{R}
\label{left_triv_der_A}
\end{IEEEeqnarray}
\item The critical points of $\Psi$, where $\mathbf{e}_{R}=0$, are $\{\mathbf{R}_{d}\}\cap\{\mathbf{R}_{d}\text{exp}(\pi S(\mathbf{s})),\mathbf{s}\in\text{S}^{2}\}$ and there exists only one critical point $\{\mathbf{R}_{d}\}$ in (\ref{L_2}).
\item $\Psi$ is locally quadratic in (\ref{L_2}), since
\begin{IEEEeqnarray}{rCl}
\lVert\mathbf{e}_{R}(\mathbf{R},\mathbf{R}_{d})\rVert^{2}\leq &\Psi(\mathbf{R},\mathbf{R}_{d})&\leq 2\lVert\mathbf{e}_{R}(\mathbf{R},\mathbf{R}_{d})\rVert^{2}
\label{quadr_Psi}
\end{IEEEeqnarray}
\end{enumerate}
\textbf{Proof of Proposition \arabic{Prop1}.} 
See \cite{qeoadapclee} for statements (\rom{1})-(\rom{5}).
See \cite{err_fun} for statements (\rom{6})-(\rom{9}).

The associated attitude error dynamics of (\ref{error_function})-(\ref{ang_vel_error}) to be used in the subsequent control design are given next.

\textbf{Proposition \arabic{Prop2}.} The error dynamics of $\{(\ref{error_function}),(\ref{att_error})\}$ satisfy:
\begin{IEEEeqnarray}{rCl}
\dot{\Psi}(\mathbf{R},\mathbf{R}_{d})&=&\mathbf{e}_{R}^{T}\mathbf{e}_{\omega}\label{dot_Psi}\\
\dot{\mathbf{e}}_{R}&=&\mathbf{E}(\mathbf{R},\mathbf{R}_{d})\mathbf{e}_{\omega}\label{der}\label{dot_Att_Error}\\
\mathbf{E}(\mathbf{R},\mathbf{R}_{d})&=&\frac{1}{2}\{ tr [ \mathbf{R}^{T}\mathbf{R}_{d} ] \mathbf{I}-\mathbf{R}^{T}\mathbf{R}_{d} \}\\
\lVert\mathbf{E}(\mathbf{R},\mathbf{R}_{d})\rVert&\leq&1\label{E_norm}\\
\lVert\dot{\mathbf{e}}_{R}\rVert&\leq&\lVert\mathbf{e}_{\omega}\rVert\label{erDot_norm}
\end{IEEEeqnarray}

Employing $\{(\ref{error_function_A}),(\ref{att_error_A})\}$, the following hold:
\begin{IEEEeqnarray}{rCl}
\dot{\Psi}(\mathbf{R},\mathbf{R}_{d})&=&\mathbf{e}_{R}^{T}\mathbf{e}_{\omega}\label{dot_Psi_A}\\
\dot{\mathbf{e}}_{R}&=&\mathbf{E}(\mathbf{R},\mathbf{R}_{d})\mathbf{e}_{\omega}\label{der}\label{dot_Att_Error_A}\\
\mathbf{E}(\mathbf{R},\mathbf{R}_{d})&=&\frac{\{ tr [ \mathbf{R}^{T}\mathbf{R}_{d} ] \mathbf{I}-\mathbf{R}^{T}\mathbf{R}_{d}+2\mathbf{e}_{R}\mathbf{e}_{R}^{T} \}}{2\sqrt{1+tr[\mathbf{R}^{T}_{d}\mathbf{R}]}}\\
\lVert\dot{\mathbf{e}}_{R}\rVert&\leq&\frac{1}{2}\lVert\mathbf{e}_{\omega}\rVert\label{erDot_norm_A}
\end{IEEEeqnarray}

The time derivative of (\ref{ang_vel_error}) is given by,
\begin{IEEEeqnarray}{rCl}
\dot{\mathbf{e}}_{\omega}&=&{}^{b}\dot{\boldsymbol{\omega}}+\mathbf{a}_{d}\IEEEnonumber\\
&=&\mathbf{J}^{-1}\left({}^{b}\mathbf{u}-{}^{b}\boldsymbol{\omega}\times\mathbf{J}{}^{b}\boldsymbol{\omega}\right)+\mathbf{a}_{d}\IEEEyesnumber\IEEEeqnarraynumspace
\label{att_error_dyn}\\
\mathbf{a}_{d}&=&S({}^{b}{\boldsymbol{\omega}})\mathbf{R}^{T}\mathbf{R}_{d}{}^{b}{\boldsymbol{\omega}}_{d}-\mathbf{R}^{T}\mathbf{R}_{d}{}^{b}{\dot{\boldsymbol{\omega}}}_{d}
\label{E_ad}
\end{IEEEeqnarray}
\textbf{Proof of Proposition \arabic{Prop2}.}
See \cite{geomquadlee_asian},\cite{qeoadapclee}, for (\ref{dot_Psi})-(\ref{erDot_norm}).
See \cite{err_fun}, for (\ref{dot_Psi_A})-(\ref{erDot_norm_A}).
See \cite{err_fun}, or \cite{geomquadlee_asian},\cite{qeoadapclee}, for (\ref{att_error_dyn})-(\ref{E_ad}).
\section{\label{appAtt}}
\textbf{Proof of Proposition \arabic{Prop3}.}
We employ a sliding methodology in (\ref{L_2}) by defining the nonlinear surface in terms of the attitude configuration errors $\{(\ref{error_function}),(\ref{att_error})\}$ or $\{(\ref{error_function_A}),(\ref{att_error_A})\}$ and apply Lyapunov analysis.
\begin{enumerate}[(a)]
\item Lyapunov candidate: We define,
\begin{IEEEeqnarray}{rCl}
V&=&\frac{1}{2k_{\omega}}\mathbf{s}_{R}^{T}\mathbf{s}_{R}+2\eta k_{R}k_{\omega}\Psi
\label{att_lyap}
\end{IEEEeqnarray}
Differentiating (\ref{att_lyap}) and substituting (\ref{att_contr}) we get,
\begin{IEEEeqnarray}{rCl}
\dot{V}&=&-\eta\mathbf{z}^{T}_{R}\mathbf{W}_{3}\mathbf{z}_{R},\mathbf{W}_{3}=\begin{bmatrix}
k_{R}^{2}&0\\
0&k_{\omega}^{2}
\end{bmatrix}
\label{Datt_lyap}
\end{IEEEeqnarray}
where $\mathbf{z}_{R}=[\lVert\mathbf{e}_{R}\rVert;\lVert\mathbf{e}_{\omega}\rVert]$.
\item Boundedness of $\Psi(\mathbf{R},\mathbf{R}_{d})$: We define the Lyapunov function,
\begin{IEEEeqnarray}{rCl}
V_{\Psi}&=&\frac{1}{2}\mathbf{e}_{\omega}^{T}\mathbf{e}_{\omega}+\eta k_{R}\Psi\label{ater_lyap}\\
\dot{V}_{\Psi}&\leq&-(\eta k_{\omega}-\frac{k_{R}}{k_{\omega}})\lVert\mathbf{e}_{\omega}\rVert^{2}\leq0
\label{dater_lyap}
\end{IEEEeqnarray}
 Equations (\ref{ater_lyap}-\ref{dater_lyap}) imply that $V_{\Psi}(t)\leq V_{\Psi}(0),\forall t\geq 0$.
Applying (\ref{surface_0}) we obtain,
\begin{IEEEeqnarray}{C}
\eta k_{R} \Psi(\mathbf{R}(t),\mathbf{R}_{d}(t)){\leq} V_{\Psi}(t){\leq} V_{\Psi}(0){<}2\eta k_{R} \label{hkrko_Psi}
\end{IEEEeqnarray}
implying that the attitude error function is bounded by,
\begin{IEEEeqnarray}{C}
\Psi(\mathbf{R}(t),\mathbf{R}_{d}(t))\leq \psi_{a} < 2,\forall t\geq 0\label{B_Psi}
\end{IEEEeqnarray}
where $\psi_{a}={V(0)}/{\eta k_{R} }$. Thus $\mathbf{R}(t)\in L_{2}$.
\item Exponential Stability: Using (\ref{low_Psi}), (\ref{upper_Psi}), for $\{(\ref{error_function}),(\ref{att_error})\}$ and (\ref{quadr_Psi}) for $\{(\ref{error_function_A}),(\ref{att_error_A})\}$ it follows that $V$ is bounded,
\begin{IEEEeqnarray}{C}
\mathbf{z}^{T}_{R}\mathbf{W}_{1}\mathbf{z}_{R}\leq V\leq\mathbf{z}^{T}_{R}\mathbf{W}_{2}\mathbf{z}_{R}
\end{IEEEeqnarray}
where $\mathbf{W}_{1}$, $\mathbf{W}_{2}$ are positive definite matrices given by,
\begin{IEEEeqnarray}{rCl}
\mathbf{W}_{1}=
\begin{bmatrix}
w_{1}&-\frac{k_{R}}{2}\\
-\frac{k_{R}}{2}&\frac{k_{\omega}}{2}
\end{bmatrix},
\mathbf{W}_{2}=
\begin{bmatrix}
w_{2}&\frac{k_{R}}{2}\\
\frac{k_{R}}{2}&\frac{k_{\omega}}{2}
\end{bmatrix}
\end{IEEEeqnarray}
\begin{IEEEeqnarray}{C}
w_{1}=\frac{k_{R}^{2}}{2k_{\omega}}+\eta k_{R}k_{\omega},w_{2}=\frac{k_{R}^{2}}{2k_{\omega}}+\frac{2}{2-\psi_{a}}\eta k_{R}k_{\omega}\label{w1w2}\\
w_{1}=\frac{k_{R}^{2}}{2k_{\omega}}+2\eta k_{R}k_{\omega},w_{2}=\frac{k_{R}^{2}}{2k_{\omega}}+4\eta k_{R}k_{\omega}\label{w1w2_A}
\end{IEEEeqnarray}
where (\ref{w1w2}) is for $\{(\ref{error_function}),(\ref{att_error})\}$ and (\ref{w1w2_A}) is for $\{(\ref{error_function_A}),(\ref{att_error_A})\}$.
Thus the following inequalities hold,
\begin{IEEEeqnarray}{C}
\lambda_{min}(\mathbf{W}_{1})\lVert \mathbf{z}_{R} \rVert^{2}\leq V \leq\lambda_{max}(\mathbf{W}_{2})\lVert\mathbf{z}_{R}\rVert^{2}\\
\dot{V} \leq -\eta\lambda_{min}(\mathbf{W}_{3})\lVert\mathbf{z}_{R}\rVert^{2}
\end{IEEEeqnarray}
Then for $\tau=\frac{\eta\lambda_{min}(\mathbf{W}_{3})}{\lambda_{max}(\mathbf{W}_{2})}$ the following holds,
\begin{IEEEeqnarray}{C}
\dot{V} \leq -\tau V
\end{IEEEeqnarray}
Thus the zero equilibrium of the attitude tracking error $\mathbf{e}_{R}$, $\mathbf{e}_{\omega}$ is exponentially stable almost globally.

Using (\ref{upper_Psi}) for $\{(\ref{error_function}),(\ref{att_error})\}$, then,
\begin{IEEEeqnarray}{C}
(2-\psi_{a})\lambda_{min}(\mathbf{W}_{1})\Psi \leq  V(t) \leq V(0)e^{-\tau t}\IEEEyesnumber
\end{IEEEeqnarray}

Using (\ref{quadr_Psi}) for $\{(\ref{error_function_A}),(\ref{att_error_A})\}$, then,
\begin{IEEEeqnarray}{C}
\frac{1}{2}\lambda_{min}(\mathbf{W}_{1})\Psi \leq  V(t) \leq V(0)e^{-\tau t}\IEEEyesnumber
\end{IEEEeqnarray}
Thus $\Psi$ exponentially decreases and from (\ref{B_Psi}) we arrive to (\ref{Psi_bou}).
This completes the proof. $\blacksquare$
\end{enumerate}

\section{\label{appB}}
\textbf{Proof of Proposition \arabic{Prop4}.} 
A sliding methodology is utilized through the definition of the surface in terms of the error vectors defined in (\ref{pos_error}), followed by Lyapunov analysis.
The position mode necessitates analysis of the coupled attitude and position dynamics.
Thus the preceding analysis of the attitude mode, is utilized here to characterize the properties of the closed loop system under the action of the controllers with the difference that $\mathbf{R}_{d}(t)$ is substituted with $\mathbf{R}_{x}(t)$.
This is because differentiation of the Lyapunov function V in (\ref{att_lyap}), parametrized by $\mathbf{R}_{x}$, gives the same result for $\dot{V}$ as in (\ref{Datt_lyap}) and thus it can be considered in (\ref{Dglo_Lyap}).
\begin{enumerate}[(a)]
\item Boundedness of $\mathbf{e}_{R}(\mathbf{R},\mathbf{R}_{x})$:
The assumptions of Proposition \arabic{Prop4} imply compliance to Proposition \arabic{Prop3} by replacing $\mathbf{R}_{d}$ with $\mathbf{R}_{x}$.
Thus the properties of (\ref{att_contr}) still apply in this analysis.
Resultantly by replacing $\mathbf{R}_{d}$ with $\mathbf{R}_{x}$, (\ref{dater_lyap}) still holds and equation (\ref{ep_0}) in (\ref{ater_lyap}) leads to,
\begin{IEEEeqnarray}{C}
\eta k_{R} \Psi(\mathbf{R}(t),\mathbf{R}_{x}(t)){\leq} V_{\Psi}(t){\leq} V_{\Psi}(0){<}\eta k_{R}\psi_{p} \label{hkrko_Psi_p}
\end{IEEEeqnarray}
signifying that the attitude error function is bounded by,
\begin{IEEEeqnarray}{C}
\Psi(\mathbf{R}(t),\mathbf{R}_{x}(t))\leq \psi_{p} < 1,\forall t\geq 0\label{Bp_Psi}
\end{IEEEeqnarray}
\item Position Error Dynamics: The analysis that follows is developed in the following domain,
\begin{IEEEeqnarray}{rCl}
D&=&\{(\mathbf{e}_{x},\mathbf{e}_{v},\mathbf{e}_{R},\mathbf{e}_{\omega})\in\mathbb{R}^{3}\times\mathbb{R}^{3}\times\mathbb{R}^{3}\times\mathbb{R}^{3}|\IEEEnonumber\\
&{ }&\Psi(\mathbf{R},\mathbf{R}_{x})\leq\psi_{p}<1\}
\label{D}
\end{IEEEeqnarray}
\textbf{Proposition \Alph{sub}.} For initial conditions in (\ref{D}), the cosine between $\mathbf{R}\mathbf{e}_{3}$ and $\mathbf{R}_{x}\mathbf{e}_{3}$ is given by $(\mathbf{R}_{x}\mathbf{e}_{3})^{T}\mathbf{R}\mathbf{e}_{3}$ and the following holds,
\begin{IEEEeqnarray}{C}
(\mathbf{R}_{x}\mathbf{e}_{3})^{T}\mathbf{R}\mathbf{e}_{3}\geq1-\Psi(\mathbf{R},\mathbf{R}_{x})>0\label{cosine}
\end{IEEEeqnarray}
The sine of the angle between $\mathbf{R}\mathbf{e}_{3}$ and $\mathbf{R}_{x}\mathbf{e}_{3}$ is given by $( (\mathbf{R}_{x}\mathbf{e}_{3})^{T}\mathbf{R}\mathbf{e}_{3})\mathbf{R}\mathbf{e}_{3}-\mathbf{R}_{x}\mathbf{e}_{3}$ and using (\ref{err_eq_psi}),
\begin{IEEEeqnarray}{C}
\lVert((\mathbf{R}_{x}\mathbf{e}_{3})^{T}\mathbf{R}\mathbf{e}_{3})\mathbf{R}\mathbf{e}_{3}-\mathbf{R}_{x}\mathbf{e}_{3}\rVert\leq\lVert\mathbf{e}_{R}\rVert\label{b_sine}
\end{IEEEeqnarray}
where for $\{(\ref{error_function}),(\ref{att_error})\}$ it holds that,
\begin{IEEEeqnarray}{C}
\lVert\mathbf{e}_{R}\rVert=\sqrt{\Psi(2-\Psi)}\leq\sqrt{\psi_{p}(2-\psi_{p})}= \theta<1\label{b_er}
\end{IEEEeqnarray}
while for $\{(\ref{error_function_A}),(\ref{att_error_A})\}$, (see \cite{err_fun}),
\begin{IEEEeqnarray}{C}
\lVert\mathbf{e}_{R}\rVert=\sqrt{\Psi(1-\frac{\Psi}{4})}\leq\sqrt{\psi_{p}(1-\frac{\psi_{p}}{4})}= \theta<1\label{b_er}
\end{IEEEeqnarray}
\textbf{Proof of Proposition \Alph{sub}.} See \cite{geomquadlee_asian},\cite{qeoadapclee}.

Equation (\ref{cosine}) is used  by adding and subtracting 
${f\mathbf{R}_{x}\mathbf{e}_{3}}{((\mathbf{R}_{x}\mathbf{e}_{3})^{T}\mathbf{R}\mathbf{e}_{3})^{-1}}$ in (\ref{eq:position}) to obtain,
\begin{IEEEeqnarray}{rCL}
m\dot{\mathbf{v}}&=&-m\frac{k_{x}}{k_{v}}\mathbf{e}_{v}-a\mathbf{s}_{x}+\mathbf{X}+m\ddot{\mathbf{x}}_{d}\IEEEyesnumber\label{eq:refposition}
\end{IEEEeqnarray}
where $f\in\mathbb{R}$, $\mathbf{X}\in\mathbb{R}^{3}$ are given by,
\begin{IEEEeqnarray}{rCL}
f&=&\lVert \mathbf{U} \rVert (\mathbf{R}_{x}\mathbf{e}_{3})^{T}\mathbf{R}\mathbf{e}_{3}\\
\mathbf{X}&=&\lVert \mathbf{U} \rVert\left( (\mathbf{R}_{x}\mathbf{e}_{3})^{T}\mathbf{R}\mathbf{e}_{3})\mathbf{R}\mathbf{e}_{3}-\mathbf{R}_{x}\mathbf{e}_{3}\right)\\
\mathbf{U}&=&mg\mathbf{E}_{3}-m\frac{k_{x}}{k_{v}}\mathbf{e}_{v}-a\mathbf{s}_{x}+m\ddot{\mathbf{x}}_{d}\label{U}
\end{IEEEeqnarray}
Then by taking the time derivative of (\ref{pos_error}), the error dynamics of $\mathbf{e}_{v}$ are given by,
\begin{IEEEeqnarray}{rCL}
m\dot{\mathbf{e}}_{v}&=&-m\frac{k_{x}}{k_{v}}\mathbf{e}_{v}-a\mathbf{s}_{x}+\mathbf{X}\label{pos_err_dyn}
\end{IEEEeqnarray}
\item Translational dynamics Lyapunov candidate: We define,
\begin{IEEEeqnarray}{rCl}
V_{x}&=&\frac{m}{2k_{v}}\mathbf{s}_{x}^{T}\mathbf{s}_{x}+ak_{x}k_{v}\mathbf{e}^{T}_{x}\mathbf{e}_{x}
\label{pos_lyap}
\end{IEEEeqnarray}
Differentiating (\ref{pos_lyap}) and substituting (\ref{pos_err_dyn}) we get,
\begin{IEEEeqnarray}{rCl}
\dot{V}_{x}&=&\mathbf{s}^{T}_{x}(-a\mathbf{s}_{x}+\mathbf{X})+2ak_{x}k_{v}\mathbf{e}^{T}_{x}\mathbf{e}_{v}\label{Dpos_lyap}
\end{IEEEeqnarray}
Using (\ref{b_sine}-\ref{b_er}), a bound of $\mathbf{X}$ is given by,
\begin{IEEEeqnarray}{rCl}
\lVert\mathbf{X}\rVert&\leq&(B+(ak_{v}+\frac{mk_{x}}{k_{v}})\lVert\mathbf{e}_{v}\rVert+ak_{x}\lVert\mathbf{e}_{x}\rVert)\lVert\mathbf{e}_{R}\rVert\IEEEnonumber\\
&\leq&(B+(ak_{v}+\frac{mk_{x}}{k_{v}})\lVert\mathbf{e}_{v}\rVert+ak_{x}\lVert\mathbf{e}_{x}\rVert)\theta\label{xbound}
\end{IEEEeqnarray}
Defining $\mathbf{z}_{x}{=}[\lVert\mathbf{e}_{x}\rVert;\lVert\mathbf{e}_{v}\rVert]$, using  (\ref{xbound}) in (\ref{Dpos_lyap}) we arrive,
\begin{IEEEeqnarray}{rCl}
\dot{V}_{x}&\leq&-\mathbf{z}^{T}_{x}\mathbf{\Pi}_{1}\mathbf{z}_{x}+\mathbf{z}^{T}_{x}\mathbf{\Pi}_{2}\mathbf{z}_{R}\label{Dpos_Lyap}
\end{IEEEeqnarray}
and by (\ref{theta}), $\mathbf{\Pi}_{1}$ is positive definite.
\item Lyapunov candidate for the complete system: We define,
\begin{IEEEeqnarray}{rCl}
V_{g}&=&V_{x}+V
\label{glo_lyap}
\end{IEEEeqnarray}
and using (\ref{low_Psi}-\ref{upper_Psi}) or (\ref{quadr_Psi}), (\ref{glo_lyap}) is bounded as follows,
\begin{IEEEeqnarray}{C}
\mathbf{z}^{T}_{R}\mathbf{W}_{1}\mathbf{z}_{R}{+}\mathbf{z}^{T}_{x}\mathbf{\Pi}_{3}\mathbf{z}_{x}{\leq}V_{g}{\leq} \mathbf{z}^{T}_{R}\mathbf{W}_{2}\mathbf{z}_{R}{+}\mathbf{z}^{T}_{x}\mathbf{\Pi}_{4}\mathbf{z}_{x}\label{glo_lyapb}\\
{\mathbf{\Pi}_{3}}{=}
{\begin{bmatrix}
ak_{x}k_{v}{+}\frac{mk_{x}^{2}}{2k_{v}}&{-}\frac{mk_{x}}{2}\\
{-}\frac{mk_{x}}{2}&\frac{mk_{v}}{2}
\end{bmatrix}}{,}\;
{\mathbf{\Pi}_{4}}{=}
{\begin{bmatrix}
ak_{x}k_{v}{+}\frac{mk_{x}^{2}}{2k_{v}}&\frac{mk_{x}}{2}\\
\frac{mk_{x}}{2}&\frac{mk_{v}}{2}
\end{bmatrix}}\IEEEnonumber
\label{glo_lyapbm}
\end{IEEEeqnarray}
and both $\mathbf{\Pi}_{3},\mathbf{\Pi}_{4}$ matrices are positive definite.
By replacing $\mathbf{R}_{d}$ with $\mathbf{R}_{x}$ in (\ref{att_lyap}) and differentiating we arrive again in (\ref{Datt_lyap}).
Using (\ref{Datt_lyap}) and (\ref{Dpos_Lyap}) the derivative of (\ref{glo_lyap}) is,
\begin{IEEEeqnarray}{rCl}
\dot{V}_{g}&\leq&-\mathbf{z}^{T}_{x}\mathbf{\Pi}_{1}\mathbf{z}_{x}+\mathbf{z}^{T}_{x}\mathbf{\Pi}_{2}\mathbf{z}_{R}-\eta\mathbf{z}^{T}_{R}\mathbf{W}_{3}\mathbf{z}_{R}\label{Dglo_Lyap}
\end{IEEEeqnarray}
\item Exponential Stability: Under the conditions (\ref{theta}-\ref{w3}) of Proposition 4 all the matrices are positive definite and for $\mathbf{z}=[\lVert\mathbf{z}_{x}\rVert;\lVert\mathbf{z}_{R}\rVert]$ equation (\ref{Dglo_Lyap}) is bounded by, 
\begin{IEEEeqnarray}{rCl}
\dot{V}_{g}&\leq&-\mathbf{z}^{T}\mathbf{\Pi}_{5}\mathbf{z},\mathbf{\Pi}_{5}{=}
\begin{bmatrix}
\lambda_{min}(\mathbf{\Pi}_{1})&-\frac{1}{2}{\lVert\mathbf{\Pi}_{2}\rVert_{2}}\\
-\frac{1}{2}{\lVert\mathbf{\Pi}_{2}\rVert_{2}}&\eta\lambda_{min}(\mathbf{W}_{3})
\end{bmatrix}\label{Dglo_Lyap1}
\end{IEEEeqnarray}
Moreover (\ref{w3}) ensures that (\ref{Dglo_Lyap1}) is negative definite.
Thus the zero equilibrium of the tracking errors of the complete system dynamics is exponentially stable in (\ref{D_x}).
A region of attraction is given by the domain (\ref{D_x}), and (\ref{ep_0}).
$\blacksquare$
\item \label{altROA}Alternative regions of exponential stability:
The Lyapunov analysis above was developed in (\ref{D_x}) without restrictions on the initial position/velocity error.
This resulted to a complicated Lyapunov analysis and a reduced region of exponential stability.
Instead if we restrict our analysis to,
\begin{IEEEeqnarray}{rCl}
D_{p}&=&\{(\mathbf{e}_{x},\mathbf{e}_{v},\mathbf{e}_{R},\mathbf{e}_{\omega})\in\mathbb{R}^{3}\times\mathbb{R}^{3}\times\mathbb{R}^{3}\times\mathbb{R}^{3}|\IEEEnonumber\\
&{ }&\Psi(0){<}\psi_{p}{<}1,\lVert\mathbf{e}_{r}\rVert<\theta,\lVert \mathbf{e}_x(0)\rVert<e_{x_{max}}\}
\label{D_xmax}
\end{IEEEeqnarray}
and bound the third order error terms that arise during the analysis using $e_{x_{max}}$ then (\ref{P}), is given by
\begin{IEEEeqnarray}{C}
\mathbf{\Pi}_{1}{=}
\begin{bmatrix}
ak_{x}^{2}(1{-}\theta)&0\\
0&ak_{v}^{2}{-}\theta(mk_{x}{+}ak_{v}^{2})
\end{bmatrix}\label{p1new}\IEEEyesnumber\\
\mathbf{\Pi}_{2}=
\begin{bmatrix}
Bk_{x}&0\\
Bk_{v}+(2ak_xk_v+\frac{mk^2_x}{k_v})e_{x_{max}}&0
\end{bmatrix}\label{p2new}
\end{IEEEeqnarray}
Alternatively a restriction on the initial velocity error results to domain,
\begin{IEEEeqnarray}{rCl}
D_{v}&=&\{(\mathbf{e}_{x},\mathbf{e}_{v},\mathbf{e}_{R},\mathbf{e}_{\omega})\in\mathbb{R}^{3}\times\mathbb{R}^{3}\times\mathbb{R}^{3}\times\mathbb{R}^{3}|\IEEEnonumber\\
&{ }&\Psi(0){<}\psi_{p}{<}1,\lVert\mathbf{e}_{r}\rVert<\theta,\lVert \mathbf{e}_v(0)\rVert<e_{v_{max}}\}
\label{D_vmax}
\end{IEEEeqnarray}
then similarly using $e_{v_{max}}$ to bound the third order error terms, $\Pi_{1}$ is given by (\ref{p1new}) and (\ref{p2new}) changes to,
\begin{IEEEeqnarray}{C}
\mathbf{\Pi}_{2}=
\begin{bmatrix}
Bk_{x}+(2ak_xk_v+\frac{mk^2_x}{k_v})e_{v_{max}}&0\\
Bk_{v}&0
\end{bmatrix}\label{p2vnew}
\end{IEEEeqnarray}
were in both cases (\ref{theta}) is given by
\begin{IEEEeqnarray}{rCl}
\theta&{<}&\min\{\frac{ak_{v}^{2}}{ak_{v}^{2}{+}mk_{x}}\}\IEEEyesnumber\label{thetaNew}\\
\IEEEnonumber
\end{IEEEeqnarray}

\squeezeup
Note that the Lyapunov analysis continues in the same manner as in Appendix \ref{appB} with (\ref{p1new}), (\ref{p2new}), (\ref{thetaNew}) (corresponding to (\ref{D_xmax})) and (\ref{p1new}), (\ref{p2vnew}), (\ref{thetaNew}) (corresponding to (\ref{D_vmax})) being utilized instead of (\ref{P}), (\ref{theta}).
It should be noted that (\ref{thetaNew}) signifies a larger basin than (\ref{theta}) but a restriction on the initial position/velocity error is introduced and this might not be desirable in some instances. 
$\blacksquare$
\end{enumerate}



\ifCLASSOPTIONcaptionsoff
  \newpage
\fi










\end{document}